\documentclass[twocolumn, trackchanges]{aastex631}
\usepackage{array}

\graphicspath{{figure/}}

\usepackage{graphicx}
\usepackage{amsmath}
\usepackage{float}
\usepackage{appendix}
\usepackage{macros}
\usepackage{chngcntr}

\newcommand{\ucdavis}{Department of Physics \& Astronomy, University of California, Davis, CA 95616, USA}

\begin{document}

\title{Inside-out versus Upside-down: The Origin and Evolution of Metallicity Radial Gradients in FIRE Simulations of Milky Way-mass Galaxies and the Essential Role of Gas Mixing}

\shorttitle{Evolution of Metallicity Radial Gradients}
\shortauthors{Graf et al.}

\correspondingauthor{Russell L. Graf}
\email{rlgraf@ucdavis.edu}

\author[0009-0009-8310-8992]{Russell L. Graf}
\affiliation{\ucdavis}

\author[0000-0003-0603-8942]{Andrew Wetzel}
\affiliation{\ucdavis}

\author[0000-0001-6380-010X]{Jeremy Bailin}
\affiliation{Department of Physics \& Astronomy, University of Alabama, Box 870324, Tuscaloosa, AL 35487-0324, USA}

\author[0000-0003-1053-3081]{Matthew E. Orr} \affiliation{Center for Computational Astrophysics, Flatiron Institute, 162 Fifth Avenue, New York, NY 10010, USA} \affiliation{Department of Physics and Astronomy, Rutgers University, 136 Frelinghuysen Road, Piscataway, NJ 08854, USA}

\begin{abstract}
Within the Milky Way (MW), younger stellar populations exhibit steeper (more negative) metallicity radial gradients; the origin of this trend remains debated.
The FIRE-2 cosmological simulations of MW-mass galaxies show the same trend as the MW, which in FIRE-2 arises because the metallicity gradient of the interstellar medium (ISM), and thus of stars at birth, became steeper over time.
We seek to understand this evolution in the context of inside-out radial growth of galaxies.
Most FIRE-2 galaxies grew radially inside-out in both gas and stars; specifically, their surface density profiles, $\Sigma(R)$, became shallower over time.
Combined with a realized superlinear (Kennicutt-Schmidt-like) relation between star formation and total gas density, the profile of the ratio $\Sigma_{\rm star}(R)/\Sigma_{\rm gas}(R)$ became shallower (flatter) over time.
Thus, if metals stayed where they were injected into the ISM from stars, the metallicity gradient would become shallower over time, as some models predict.
However, metallicity gradients in FIRE-2 became steeper over time, because of the additional effects of (radial) mixing of metals in the ISM.
Specifically, the velocity dispersion and net radial advection of gas declined over time, as ISM turbulence decreased and the disk settled, leading to upside-down vertical growth.
In FIRE-2, this evolution in metal mixing of gas associated with upside-down growth dominates over inside-out radial growth, causing the metallicity radial gradient of the ISM and of stars at birth to become steeper over time.
We argue that this reflects the ISM history of the MW and of typical MW-mass galaxies.
\end{abstract}

\keywords{}

\section{Introduction}
\label{sec:intro}

Metals across a galaxy's interstellar medium (ISM) and stars can form spatial patterns that reflect the galaxy's formation history.
In many cases, the most significant source of spatial variation is a radial gradient.
\citet{Searle1971} demonstrated that galaxies have negative radial gradients in their ISM metallicity, such that gas at a smaller galactocentric radius, $R$, is more metal-rich, on average.
Such negative metallicity gradients are common, and many works have quantified them in the Milky Way (MW), M31, and nearby galaxies, both in the ISM and in newly-born stars \citep[for example,][]{Sanders2012, SanchezMenguiano2016, Belfiore2017, Sakhibov2018, Kreckel2019, Kreckel2020, Wenger2019, Molla2019, Zinchenko2019, Hernandez2021, Hawkins2023}.
As we discuss further in Section~\ref{sec:background}, a galaxy, in principle, should develop a negative metallicity radial gradient in its ISM if its density profile in stars, which enrich the metals, is steeper than that of its gas, which dilutes the metals \citep[for example,][]{Chiappini2001, Ho2015}.

Though negative radial gradients in ISM metallicity are common in the local Universe, there is little consensus, observationally or theoretically, for how metallicity gradients in the ISM evolve across time: whether they generally become steeper, shallower, or remain fixed across the formation history of a galaxy.
Historically, observations of galaxies at high-redshift offer mixed results, with some showing negative gradients \citep[for example,][]{Jones2013}, others showing flat gradients \citep[for example,][]{Curti2020}, and others showing positive gradients \citep[for example,][]{Wang2022}.
Such observations have been difficult, requiring spatially-resolved spectroscopy of faint (distant) galaxies, leading to large uncertainties with weak constraints on theoretical models \citep[see for example Figure~7 of][]{Hemler2021}.
Promisingly, the James Webb Space Telescope (JWST) now provides a new window into this regime: \citet{Ju2024} analyzed a sample of 26 star-forming galaxies at $z \sim 1$ and found that they generally have relatively flat radial gradients in ISM metallicity, with steeper (more negative) gradients in more massive, ``diskier'' galaxies.

A complementary approach to understanding how metallicity gradients evolve is via the information encoded within an individual galaxy like the MW.
While the metallicity patterns in the ISM changed over time via processes like enrichment, dilution, and mixing, individual stars essentially retain the (atmospheric) metallicity with which they were born.
Because stars reflect the ISM from which they formed, one can infer the history of the ISM via stellar populations today, through an approach sometimes known as ``chemical tagging'' \citep{FreemanBlandHawthron2002}. 

Many works characterized metallicity radial gradients for stellar populations versus their age in the MW \citep[for example,][]{Vickers2021, Lian2022, Lian2023, Willet2023, Anders2023, Imig2023}.
These works consistently found that metallicity radial gradients are strongest for the youngest stars and weaken with increasing age.
The simplest interpretation is that the metallicity radial gradient in the MW's (star-forming) ISM became steeper over time, as some works argue \citep[for example,][]{Chiappini2001, Ma2017a, Ma2017b, VK2020, Sharda2021, Bellardini2021}.

However, a key caveat to interpreting the results of stellar chemical tagging is that stars can move from their birth radii via radial redistribution \citep[for example,][]{SellwoodBinney2002, Haywood2008, Okalidis2022}.
This generally weakens the radial gradient for a given population of stars, and because older stars have had more time to redistribute, a common expectation is that older stellar populations have experienced more flattening of their gradient \citep[for example,][]{Minchev2013}.
Therefore, many works instead interpret the flattening of the metallicity gradient with stellar age in the MW as primarily arising from this age dependence of stellar radial redistribution, and that the MW's ISM metallicity gradient was in fact steeper (or similarly steep) in the past and became shallower (flatter) over time because of inside-out radial growth \citep[for example,][]{Minchev2018, VK2018, Agertz2021, Buck2023}, as we describe in Section~\ref{sec:background}.

Cosmological simulations show near universal consensus that galaxies grow radially inside-out \citep[for example,][]{Brook2012, Bird2013, Ma2017b, Buck2020, Iza2024}.
Observational evidence of inside-out radial growth, by comparing the expected progenitors of MW-mass galaxies at different redshifts, is mixed: for example, \citet{Pan2023} and \citet{Mun2024} found inside-out growth, while \citet{Hasheminia2022} and \citet{Tan2024} did not, and \citet{Hasheminia2024} found that only galaxies with $\Mstar \gtrsim 10^{10.7} \Msun$ today show inside-out radial growth.
That said, observational analyses of the MW consistently find strong evidence of inside-out radial growth in its history \citep[for example,][]{Frankel2019, Anders2023, Lian2024, Wang2024}.

Thus, theoretical works nearly universally agree that galaxies grow radially inside-out, but theoretical expectations for how the ISM metallicity radial gradient evolves in a galaxy are highly mixed, and this remains one of the most controversial topics among models of galaxy evolution.
The simulations of 10 MW-mass galaxies in \citet{VK2018}, the NIHAO-UHD simulations of 4 MW-mass galaxies \citep{Buck2023}, the FOGGIE simulations of 6 MW-mass galaxies \citep{Acharyya2024}, and the MW model of \citet{Minchev2018}, for example, show ISM metallicity radial gradients that began steep (negative) and flattened over time.
On the other hand, the simulation of a single MW-mass galaxy in \citet{VK2020}, some of the NIHAO-UHD simulations as presented in \citet{Lu2022b}, the MW models of \citet{Chiappini2001} and \citet{Sharda2021}, and the VINTERGATAN simulation of a single MW-mass galaxy \citep{Renaud2024}, for example, show metallicity gradients that began flat/shallow and steepened (became more negative) over time, with the caveat that while VINTERGATAN shows steepening gradients in young stars, it shows the opposite trend in its overall ISM \citep{Agertz2021}.
The FIRE-2 cosmological simulations, including analyses of 11 MW-mass galaxies, show metallicity gradients that strengthen over time while the galaxies grow radially inside-out \citep{Ma2017a, Ma2017b, Bellardini2021, Carillo2023}.
Thus, many of these results indicate that inside-out radial growth does not necessarily cause metallicity gradients to become shallower (flatter) over time.

Mixing is another key factor that regulates the spatial distribution of metals in the ISM of a galaxy.
Processes such as turbulence, feedback, accretion, mergers, and satellite interactions can lead to the mixing of metals in the ISM that generally tend to flatten radial gradients, especially at early times when the turbulence was higher, so the ISM was geometrically thicker \citep[for example,][]{Bird2012, Bird2021, McCluskey2023, Han2024}.
Related processes include galactic winds, which can eject metals out of the galaxy \citep[for example,][]{Hayward2017, Jones2018, Pandya2021}, and radial advection, which can move pristine gas and/or metal-rich gas in bulk flows across the disk \citep[for example,][]{Sharda2021}.
For example, \citet{Orr2023} found that, at late cosmic times, spiral arms can cause significant azimuthally-dependent bulk radial advection in the gas, and, in the presence of a metallicity radial gradient, they argued that this sources most of the azimuthal variations in gas metallicity today.
Critical for the evolution of metallicity gradients, the strengths of these processes evolve over time.
Indeed, works like \citet{Bird2013} argued that MW-mass galaxies evolved radially ``inside-out'' but vertically ``upside-down'', because higher gas turbulence at earlier times led to a dynamically hotter and geometrically thicker ISM, which settled over time.

In this work, we use the FIRE-2 cosmological simulations to study the origin of the evolution of metallicity radial gradients across the histories of MW-mass galaxies.
We build on the FIRE-2 analyses of \citet{Bellardini2021} and \citet{Bellardini2022}, who showed that the metallicity radial gradient in the ISM and in stars at birth was nearly flat, on average, at early times and became steeper (more negative) over time, being largely consistent with measurements of nearby galaxies today.
We also build on the FIRE-2 analysis of \citet{Graf2024}, who showed that these same trends persist for stars today as a function of their age, such that the trend for stars at birth and stars today are similar, with only a mild shallowing from radial redistribution of stars after birth.
\citet{Graf2024} also showed that this trend in FIRE-2 for stars today is qualitatively similar to trends observed in the MW, providing evidence that the ISM metallicity radial gradient in the MW began flatter and strengthened (steepened) over time.

\section{Methods}
\label{sec:methods}

\subsection{FIRE-2 simulations}
\label{sub:FIRE2}

We analyze 11 galaxies with masses similar to the MW (and M31) from the FIRE-2 cosmological zoom-in simulations \citep{Hopkins2018}, all of which are publicly available \citep{Wetzel2023}.
Of these $11$ galaxies, $5$ are cosmologically isolated galaxies from the \textit{Latte} suite \citep[introduced in][]{Wetzel2016}: m12b, m12c, m12f, m12i, and m12m. These have halo masses of $1 \! - \! 2 \times 10^{12} \Msun$, dark-matter mass resolution of $\approx 3.5 \times 10^{5} \Msun$, and initial baryonic particle masses of $7070 \Msun$ (because of stellar mass loss, most star particles have $\approx 5000 \Msun$).
The other $6$ galaxies are from the ELVIS (Exploring the Local Volume in Simulations) FIRE suite, which consists of 3 Local Group-like galaxy pairs: Romeo $\&$ Juliet, Romulus $\&$ Remus, and Thelma $\&$ Louise.
Their mass resolution is $\approx 2$ times better than \textit{Latte}, with initial baryonic particle masses of $3500 \! - \! 4000 \Msun$.

Star and dark-matter particles have fixed gravitational softenings, with Plummer equivalents of $\epsilon_{\rm star} = 4 \pc$ and $\epsilon_{\rm DM} = 40 \pc$, comoving at $z > 9$ and physical at $z < 9$.
Gas cells have fully adaptive gravitational softening, which matches the hydrodynamic kernel smoothing, and reaches a minimum of $1$ pc; in the typical ISM at densities $\approx 1 \cci$, the smoothing is $\approx 40 \pc$.

To generate initial conditions, we used MUSIC \citep{HahnAbel2011}. Each zoom-in region is embedded within a cosmological box with side lengths ranging from $70.4 - 172 \Mpc$. These simulations assume a flat $\Lambda$CDM cosmology with parameters that broadly agree with \citep{PlanckCollab2020}: $h = 0.68 - 0.71$, $\Omega_\Lambda = 0.69 - 0.734$, $\Omega_{\rm m} = 0.266 - 0.31$, $\Omega_{\rm b} = 0.0455 - 0.048$, $\sigma_{\rm s} = 0.801 - 0.82$, and $n_{\rm s} = 0.961 - 0.97$. Each simulation stores $600$ snapshots spanning $z = 99$ to $z = 0$, each with time spacing $\lesssim 25 \Myr$.

We ran these simulations using the \textsc{Gizmo} code \citep{Hopkins2015}, with the mesh-free finite-mass (MFM) mode for hydrodynamics, which is a quasi-Lagrangian Godunov method that provides adaptive spatial resolution while maintaining exact mass, energy, and momentum conservation, excellent conservation of angular momentum, and accurate capturing of shocks. For gravity, \textsc{Gizmo} uses an updated version of the Tree-PM solver from GADGET-2 \citep{Springel2005}.

The simulations use the FIRE-2 physics model \citep{Hopkins2018}, which includes a redshift-dependent and spatially-uniform cosmic ultraviolet background from \citet{FaucherGiguere2009}. They also include metallicity-dependent radiative heating and cooling processes for gas across $10 \! - \! 10^{10}$ K, including: photoionization and recombination, free-free, Compton, photoelectric and dust collisional, cosmic ray, metal-line, molecular, and fine-structure processes, accounting for $11$ elements (see below).

A gas cell turns into a star particle on a local gravitational free-fall time when it becomes self-gravitating, Jeans-unstable, cold ($T < 10^4$ K), and molecular \citep[following][]{KrumholzGnedin2011}.
A star particle inherits the mass and metallicity of its progenitor gas cell, representing a mono-age, mono-abundance stellar population, assuming a \citet{Kroupa2001} initial mass function.

\begin{table*}[!htb]
\begin{center}
\caption{
Properties today of the 11 FIRE-2 galaxies we analyze
}
\begin{tabular*}{0.581 \linewidth}{@{}c|ccccccc@{}}
\hline
galaxy & $M_\star^{\rm 90}$ & $R_\star^{\rm 90}$ & $R_{\rm \star,young}^{\rm 90}$ & $R_\star^{\rm 50}$ & $R_{\rm \star,young}^{50}$ & $\frac{\partial {\rm [Fe/H]}}{\partial R}_{\rm \star,young}$ \\
name & [$10^{10} \Msun$] & [kpc] & [kpc] & [kpc] & [kpc] & [dex kpc$^{-1}$]\\
\hline
m12m$^{1}$ & $10.0$ & $11.0$ & $13.4$ &  $5.5$ & $6.7$ & $-0.036$\\
Romulus$^{2}$ & $8.0$ & $14.2$ & $17.0$ & $3.9$ & $9.0$  & $-0.030$\\
m12b$^{3}$ & $7.3$ & $8.8$ & $10.7$ & $2.1$ & $5.0$ & $-0.042$\\
m12f$^{4}$ & $6.9$ & $13.9$ & $18.8$ & $3.5$ & $10.5$ & $-0.027$\\
Thelma$^{3}$ & $6.3$ & $10.7$ & $16.0$ & $4.1$ & $8.4$ & $-0.026$\\
Romeo$^{3}$ & $5.9$ & $13.5$ & $23.8$ & $4.4$ & $9.9$ &  $-0.032$\\
m12i$^{5}$ & $5.3$ & $9.3$ & $13.7$ & $3.2$ & $6.9$ & $-0.034$\\
m12c$^{3}$ & $5.1$ & $8.6$ & $10.5$ & $3.3$ & $4.4$ & $-0.039$\\
Remus$^{2}$ & $4.0$ & $11.8$ & $21.1$ & $3.3$ & $7.5$  & $-0.027$\\
Juliet$^{3}$ & $3.3$ & $8.1$ & $19.4$ & $2.3$ & $7.9$ & $-0.036$ \\
Louise$^{3}$ & $2.3$ & $11.8$ & $22.9$ & $3.1$ & $10.8$ & $-0.021$ \\
\hline
mean & $5.9$ & $11.1$ & $17.0$ & $3.5$ & $7.9$ & $-0.031$ \\
\hline
\end{tabular*}
\label{tab:galaxy_properties}
\tablecomments{
Galaxies are in decreasing order of stellar mass today.
Here, ``young'' refers to stars with ages $< 100 \Myr$.
Columns show: galaxy name; $M_\star^{\rm 90}$ is the stellar mass within $R_\star^{\rm 90}$; $R_\star^{\rm 90}$ is the radius that encloses $90\%$ of the mass of all stars within $30 \kpc$; $R_{\rm \star,young}^{\rm 90}$ is the radius that encloses $90\%$ of the mass of \textit{young} stars; $R_\star^{\rm 50}$ is the radius that encloses $50\%$ of the stellar mass within $30 \kpc$; $R_{\rm \star,young}^{\rm 50}$ is the radius that encloses $50\%$ of mass of \textit{young} stars; the radial gradient of [Fe/H] for \textit{young} stars, across $0 - R^{90}_{\rm \star, young}$.
The publication that introduced each simulation at this resolution is: \citet{Hopkins2018}$^{1}$, \citet{GarrisonKimmel2019a}$^{2}$, \citet{GarrisonKimmel2019b}$^{3}$, \citet{GarrisonKimmel2017}$^{4}$, \citet{Wetzel2016}$^{5}$.
}
\end{center}
\end{table*}

We analyze only galaxies with stellar masses within a factor of $\approx 2$ of the MW \citep[for example,][]{BlandHawthornGerhard2016}. Table~\ref{tab:galaxy_properties} shows key properties today of these 11 MW-mass FIRE-2 galaxies.
These galaxies experienced significant variation in their formation histories \citep{Santistevan2020, McCluskey2023}, because their selection was indifferent to any specific properties beyond halo mass (with the additional constraint of a LG-like paired-halo environment for the ELVIS suite).
For most results, we show the mean and standard deviation (galaxy-to-galaxy scatter) across these 11 MW-mass galaxies.
Therefore, our results should represent trends and formation histories typical of MW-mass galaxies today.

\subsection{Metal enrichment in FIRE-2 simulations}
\label{sub:metal_enrichment_in_FIRE2}

FIRE-2 models key stellar evolution processes, including stellar winds, core-collapse and white-dwarf (Ia) supernovae, radiation pressure, photoionization, and photoelectric heating. For stellar wind rates and nucleosynthetic yields, FIRE-2 uses a combination of models \citep{VandenHoek1997, Marigo2001, Izzard2004} synthesized in \citet{Wiersma2009}. FIRE-2 uses core-collapse supernova (CCSN) rates from STARBURST99 \citet{Leitherer1999} and yields from \citet{Nomoto2006}, and white-dwarf supernova (WDSN) rates from \citet{Mannucci2006} with nucleosynthetic yields from \citet{Iwamoto1999}.
FIRE-2 tracks 11 elements (H, He, C, N, O, Ne, Mg, Si, S, Ca, Fe).
We scale elemental abundances to (proto)Solar values from \citet{Asplund2009}.

We focus our analysis on radial gradients in [Fe/H], because Fe is the most commonly measured metal in stars.
We checked that our results hold true for the other metals above, including O (more commonly measured in the ISM), and as \citet[Appendix B of][]{Graf2024} showed, the radial gradients in other elements show the same trends as for [Fe/H]; the only key difference is an offset in the exact strength of the gradient, at the level of $\approx 0.005$ dex kpc$^{-1}$.
In FIRE-2, CCSN and WDSN generate Fe.
All CCSN occur within the first $\approx 38 \Myr$ of the birth of a stellar population, and given the assumed rates from \citet{Mannucci2006}, about half of all WDSN occur $\approx 38 - 100 \Myr$ after star formation (see Appendix~A of \citealt{Hopkins2018} and Figures~2 and 4 of \citealt{Hopkins2023}).
As a result, the majority ($\approx 2/3$) of all Fe enrichment in FIRE-2 occurs within the first $\approx 100 \Myr$ of a given stellar population.
Thus, $100 \Myr$ is a characteristic timescale that motivates our definition of a ``young'' stellar population below.

Also important, FIRE-2 explicitly models the subgrid mixing and diffusion of metals via unresolved turbulent eddies \citep{Su2017, Escala2018, Hopkins2018}.
\citet{Bellardini2021} showed that the details of this diffusion model do not significantly affect the radial or vertical gradients of metallicity in FIRE-2 MW-mass galaxies, but they do affect the azimuthal scatter (at fixed radius).
This implies that most of the mixing of gas relevant for galaxy-wide (azimuthally-averaged) radial gradients occurs on \textit{resolved} scales in these FIRE-2 simulations.
\citet{Bellardini2021} also showed that these FIRE-2 MW-mass galaxies have radial gradients and azimuthal scatter of metals in the ISM that are similar to those observed in nearby galaxies of similar mass.
That said, \citet{Bellardini2022} and \citet{Graf2024} showed that the metallicity radial gradient in stars is generally weaker than the MW (see also Figure~\ref{fig:mw_compare}).

\subsection{Selecting and measuring stars}
\label{sub:selecting_and_measuring_stars}

At each snapshot, we identify each galaxy's center using an iterative zoom-in on all star particles that end up in the galaxy today \citep[see][]{Wetzel2023}.
We then identify the orientation of the disk at each snapshot by measuring the moment-of-inertia tensor using the $25\%$ youngest stars within $\approx 10 \kpc$ of the galaxy center.
In our analysis of radial profiles and gradients, we include only stars ``in the disk", within a height of $|Z| < 3 \kpc$ of the disk midplane, though our results are not sensitive to the details of this threshold.
When evaluating any property for each galaxy, we measure the median across the stellar population at a given time and spatial location.
We then compute the mean and standard deviation across all 11 galaxies.

To determine the size of a galaxy at a given lookback time, $t_{\rm lb}$, we calculate $R_\star^{\rm 50}(t_{\rm lb})$ and $R_\star^{\rm 90}(t_{\rm lb})$, the radii that enclose $50\%$ and $90\%$ of the mass of all stars within $30 \kpc$ comoving \citep[following][]{Bellardini2022}.
We also compute these radii using only ``young'' stars, with ages $< 100 \Myr$ at that time.
Because our motivation is to understand the metallicity gradient in the ISM \textit{as it relates to the formation of stars}, we measure metallicity gradients of young stars and gas out to $R_{\rm \star,young}^{\rm 90}(t_{\rm lb})$.
We measure radial gradients by fitting a single linear relation across the radial range using numpy.polyfit.
As \citet{Bellardini2022} showed (and we show below), the metallicity profiles are not always perfectly linear; consistent with the MW, they are somewhat steeper at smaller radii. We discuss this subtlety more below, but it does not affect any of our qualitative results.

Throughout, we measure metallicity gradients in physical units, dex kpc$^{-1}$, rather than scaled in units of dex $R_{90}^{-1}$.
Our choice is motivated by the tests in \citet{Bellardini2021} and \citet{Graf2024}: analyzing the same FIRE-2 galaxies, they found the galaxy-to-galaxy scatter in [Fe/H] radial and vertical gradients is smallest (most self-similar) in physical units, rather than in units scaled to each galaxy's radius.
Furthermore, we checked that none of our results here depend qualitatively on these details regarding how we measure the gradient.

\section{Expectations for the evolution of metallicity radial gradients}
\label{sec:background}

For theoretical background, we first reiterate arguments for why a galaxy's ISM should form a negative radial gradient in metallicity \citep[for example,][]{Ho2015, Ma2017a}, and we present arguments for how the gradient could evolve.

Stars generate metals, and young stellar populations, in particular, source most of the \textit{recent} metal enrichment.
Recent enrichment, which has not yet had enough time to mix and homogenize across the ISM, is mostly likely to contribute to metallicity spatial variations in the ISM.
Therefore, the total mass in metals should scale with the total mass in stars formed, and the recent enrichment in metals should scale with the recent star formation and thus mass of ``young'' stars.
Using Fe as a fiducial metal, one conventionally cites ``metallicity'' as [Fe/H], that is, the total mass in Fe relative to H (scaled to the Solar value).
H always dominates the overall gas mass, on galactic scales.
Therefore, to \textit{source} a metallicity gradient, the mass in stars (up to some age threshold), relative to the mass in gas (over which the metals are diluted), should depend on galactocentric radius, $R$.
Approximately then,
\begin{equation}
{\rm [Fe/H]}(R) \approx \log\left( \frac{\Sigma_{\rm \star,young}(R)}{ \Sigma_{\rm gas}(R)} \right) + C
\label{eq:profile}
\end{equation}
\begin{equation}
\frac{ \partial {\rm [Fe/H]} }{ \partial R } \approx \frac{ \partial \log\left( \Sigma_{\rm \star,young} / \Sigma_{\rm gas} \right) }{ \partial R }
\label{eq:gradient}
\end{equation}
where $\Sigma_{\rm \star,young}(R)$ is the mass surface density in ``young'' stars (which we define and discuss more below), $\Sigma_{\rm gas}(R)$ is the mass surface density in total gas, and $C$ is a constant.

\textit{In this framing, $\Sigma_{\rm gas}(R)$ represents the instantaneous density profile of the gas, which has already been shaped by processes like dilution from recent cosmic accretion}.
Furthermore, measuring $\Sigma_{\rm \star,young}(R)$ at a single instant assumes that stars did not move significantly in $R$ as they enriched the ISM.

One can make the strong assumption that metals stayed in the ISM where they were injected from stars.
In this simple limit, if $\Sigma_{\rm \star,young}(R) / \Sigma_{\rm gas}(R)$ declines with $R$, then the galaxy's ISM should form a negative metallicity gradient.
This decline with $R$ should generally occur from a combination of two observed effects.
First, the surface density profile of gas in a galaxy generally declines with $R$, approximately as $\Sigma_{\rm gas} \propto \exp(-R / R_{\rm s})$, where $R_{\rm s}$ is the scale radius \citep[for example,][]{SaintongeCatinella2022}.
Second, the star-formation rate generally increases superlinearly with the \textit{total} gas density on kpc scales \citep{Schmidt1959, Kennicutt1998}, such that $\Sigma_{\rm SFR} \propto \Sigma_{\rm gas}^{n}$, typically with $n \approx 1.4$, although in this case the only requirement is $n > 1$.
Thus, if $\Sigma_{\rm gas}$ decreases with $R$, $\Sigma_{\rm \star,young}$ decreases more rapidly with $R$, so the ratio $\Sigma_{\rm \star,young}(R) / \Sigma_{\rm gas}(R)$ declines with $R$.
Over a relevant metal-enrichment timescale, Equations~\ref{eq:profile} and \ref{eq:gradient} show that this would lead to a negative radial profile/gradient in ISM metallicity, if metals stayed where they were injected.
Indeed, \citet{Bellardini2022} showed that for MW-mass galaxies in FIRE-2 simulations, Equation~\ref{eq:profile} holds to good approximation today, though they did not explore this behavior across time, as we do here.

In this framework, how would the metallicity radial gradient in the ISM evolve across the history of a galaxy?
As discussed in Section~\ref{sec:intro}, cosmological simulations agree that galaxies generally grow radially inside-out, whereby greater fractional growth occurs at larger $R$ (relative to smaller $R$) over time.
Specifically, cosmological accretion of gas tends to occur preferentially at larger $R$ over time, which leads $\Sigma_{\rm gas}(R)$ to become shallower over time. Because star formation increases superlinearly with total gas density, $\Sigma_{\rm \star,young}(R)$ becomes shallower more rapidly than $\Sigma_{\rm gas}(R)$, so the right-hand sides of Equations~\ref{eq:profile} and \ref{eq:gradient} become shallower over time.
This leads to a common expectation that inside-out radial growth of a galaxy should cause its ISM metallicity radial gradient to become shallower (flatter) over time \citep[for example,][]{Minchev2018}.

Again, critical to this argument is the strong assumption that metals stay at the same $R$ where stars injected them into the ISM, that is, that metal mixing (or loss) is weak.
Therefore, one must additionally consider the criterion for a galaxy to \textit{sustain} a metallicity gradient: the ISM must not experience too much mixing of metals (via mechanisms like turbulence, azimuthally-dependent bulk flows, patchy star-formation) or radially-dependent loss of metals, via metal-enriched outflows from galactic winds.
Thus, Equations~\ref{eq:profile} and \ref{eq:gradient} effectively assume, among other things, that the ISM is in a well-ordered, rotationally-supported disk with little radial motion of gas, that is, with $v_{\rm \phi} / \sigma_{\rm v} \gg 1$, where $v_{\rm \phi}$ is the rotational velocity and $\sigma_{\rm v}$ is the \textit{azimuthally-averaged} (not just local) velocity dispersion, and these represent the dynamics of the \textit{total} (not just the cold, star-forming) ISM, given that metals are injected into and mix between different phases of the ISM.

Many works have shown that the strength of these processes, which drive metal mixing, vary across the histories of MW-mass galaxies.
Specifically, over time a MW-mass galaxy declines in: gas accretion rates and internal radial advection \citep[for example,][]{Sharda2021, Trapp2022}, rates and burstiness of star formation \citep[for example,][]{Yu2023}, velocity dispersion, turbulence, and disk thickness \citep[for example,][]{Bird2021, McCluskey2023}, and outflows from galactic winds \citep[for example,][]{Pandya2021, Orr2022a, Orr2022b}.
Indeed, this decline in ISM turbulence over time leads to the expectation that disk galaxies grow radially ``inside-out'' but vertically ``upside-down'' \citep{Bird2013}.
Observations support the above trends, including the decline of gas turbulence towards lower redshifts \citep[for example,][]{Kassin2012, Wisnioski2015, Ubler2019, Rizzo2020}, and that more than half of metals can be lost from the ISM in outflows \citep{Jones2018}.
Even if outflow metals reaccrete into the ISM, if they do so at a different $R$ this effectively acts further to mix the metals across the galaxy.
The decline in metal mixing in gas over time, as the disk settled and $v_{\rm \phi} / \sigma_{\rm v}$ increased, would act to make the metallicity gradient steeper over time, and this could create a correlation between the strength of the metallicity radial gradient and $v_{\rm \phi} / \sigma_{\rm v}$ \citep[for example,][]{Ma2017a, Bellardini2022}.

Therefore, in terms of expectations for the evolution of the ISM metallicity radial gradient, inside-out radial growth implies a shallowing (flattening) over time, but disk settling and upside-down vertical growth, which leads to a reduction in metal mixing, implies a strengthening (steepening) over time.
\textit{Because these two processes, inside-out versus upside-down growth, drive the metallicity radial gradient in opposite directions, the realized evolution of a galaxy will depend on which one dominates, and at which times.}

\section{Results}
\label{sec:results}

In the following subsections, we first present the evolution of the metallicity radial gradient in the FIRE-2 simulations of MW-mass galaxies, showing that their ISM gradients became steeper (more negative) over time, and that the trend this imparts in stars is broadly consistent with the trend observed for stars today in the MW.
We then show that these FIRE-2 galaxies grow radially inside-out.
Finally, we argue that their metallicity gradients became steeper over time despite their inside-out growth primarily because of the strong evolution of the mixing of metals in the ISM.

\subsection{Evolution of metallicity radial gradient: comparison with the Milky Way}
\label{sub:mw_compare}

\begin{figure}
\centering
\includegraphics[width = \columnwidth]{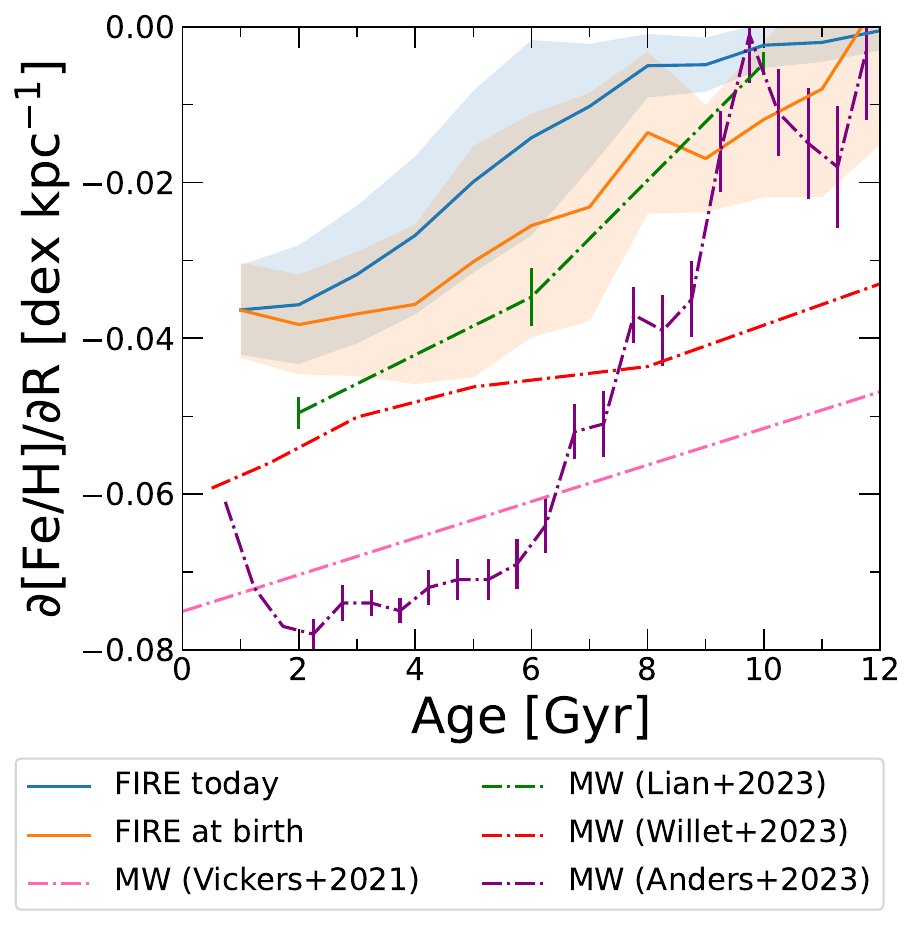}
\caption{
Radial gradient of metallicity, via $\partial {\rm [Fe/H]} / \partial R$, for stars today versus their age, and for the same stars at birth, adapted from \citet{Graf2024}.
For each FIRE-2 galaxy, we measure the gradient across $R = 0 - R^{\rm 90}_{\rm \star,young}(t_{\rm lb})$, where we measure $R^{\rm 90}_{\rm \star,young}$ using young stars ($< 100 \Myr$) at each age (lookback time), $t_{\rm lb}$.
Each solid line and shaded region shows the mean and standard deviation across 11 FIRE-2 galaxies.
The radial gradient is almost always negative, both at birth and today, becoming steeper for stars that formed more recently, and nearly flat for stars that formed $\gtrsim 10 \Gyr$ ago. 
Because of radial redistribution after birth, the gradient is marginally shallower today than at birth, but the difference is typically $\lesssim 0.01$ dex kpc$^{-1}$.
We also show observations of the MW from \citet{Vickers2021}, \citet{Lian2023}, \citet{Willet2023}, and \citet{Anders2023}.
FIRE-2 simulations have shallower gradients than the MW at most ages, though the steepest gradients in FIRE-2 are broadly consistent with \citet{Lian2023}.
\textit{Importantly, FIRE-2 agrees with all of these MW observations that the gradient is shallower (flatter) for older stars, and with \citet{Lian2023} and \citet{Anders2023} that it is nearly flat for the oldest stars.
In FIRE-2 this primarily reflects the metallicity gradient of stars at birth, that is, of the star-forming ISM.}
}
\label{fig:mw_compare}
\end{figure}

Figure~\ref{fig:mw_compare} \citep[adapted from Figure~3 in][]{Graf2024} shows the radial gradient of [Fe/H] in stars, $\partial {\rm [Fe/H]} / \partial R$, versus stellar age, in the FIRE-2 galaxies and in various observational analyzes of the MW.
We show the radial gradient in FIRE-2 using the galactocentric radii of stars both today and at birth, which we measure across $R = 0 - R_{\rm \star,young}^{\rm 90}(t_{\rm lb})$ at each lookback time, $t_{\rm lb}$.
Given the snapshot time spacing in FIRE-2, ``at birth'' means measuring a star particle's radius within $\lesssim 25 \Myr$ of when it formed.
As \citet{Graf2024} found, the radial gradient steepens for younger stars, both today and at birth.
The gradient was modestly ($\approx 0.01$ dex kpc$^{-1}$) steeper at birth than today, given the effects of radial redistribution after birth (Bellardini et al. in prep.).
However, the trends are similar.
Importantly, this means that this trend for stars today was set primarily by the radial gradient of stars at birth, which in turn was set by the radial gradient in the (star-forming) ISM.

Figure~\ref{fig:mw_compare} also compares 4 observational analyses of stars in the MW: \citet{Vickers2021}, \citet{Lian2023}, \citet{Willet2023}, and \citet{Anders2023}.
See \citet{Graf2024} for a full description of these works.
The radial gradient in the MW is steeper than in FIRE-2 at most ages, ranging between $\approx -0.05$ dex$\kpc^{-1}$ and $-0.075$ dex$\kpc^{-1}$ for the youngest stars, whereas the steepest gradient in FIRE-2 today is $-0.045$ dex$\kpc^{-1}$. That being said, Figure~\ref{fig:mw_compare} shows that the results of \citet{Lian2023} lie within $1 \sigma$ of FIRE-2 for most stellar ages. The main relevant trend for this work, however, is that all works agree that the gradient is generally shallower for older stars.

These results motivate our analysis in this work. Specifically, the FIRE-2 simulations show the same trend with age today as these MW observations: the metallicity radial gradient is steeper (more negative) for younger stars.
In FIRE-2 this trend emerges primarily because of the evolution of stars at birth, that is, of the star-forming ISM.
In particular, in FIRE-2, radial redistribution of stars after birth does not qualitatively change/reverse this trend with age, beyond modestly shallowing the gradient at all ages.
Therefore, our analysis below focuses on understanding the origin of this trend in the star-forming ISM (effectively, in stars at birth) in FIRE-2, including how it relates to the inside-out radial growth of these galaxies and the evolution of metal mixing in gas.

These results for FIRE-2 in Figure~\ref{fig:mw_compare} contrast with expectations/assumptions from other works (such as \citealt{Minchev2018, Anders2023, Ratcliffe2023, Lian2023}), that the ISM radial gradient in the MW was steeper at earlier times, and that radial redistribution of stars since their birth has flattened the gradient as measured today.
Those works argue that, because the amount of stellar radial redistribution increases with time, it inverts the apparent evolution of the gradient as reflected in stars versus their age today.
As we argue below, one must also consider the radial redistribution (mixing) of gas, especially at early times.

\subsection{Inside-out radial growth}
\label{sub:inside_out_growth}

We first quantify the radial growth of these $11$ MW-mass galaxies, to provide context for understanding the evolution of metallicity radial gradients, building on previous work that has demonstrated inside-out radial growth for (some of) these FIRE-2 galaxies \citep{Ma2017b, Bellardini2022, Carillo2023}.

\begin{figure}
\centering
\includegraphics[width = \columnwidth]{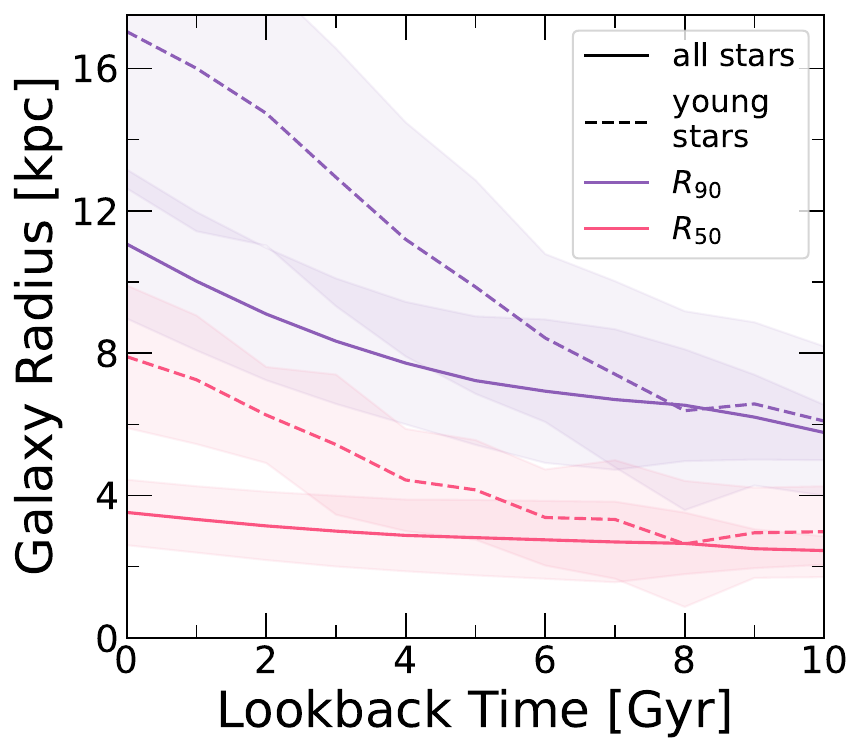}
\vspace{-5 mm}
\caption{
A metric of ``inside-out" radial growth: the cylindrical radius that encloses $50\%$ or $90\%$ of the stellar mass within an aperture of $30 \kpc$ comoving, $R_{\star}^{50}(t_{\rm lb})$ or $R_\star^{\rm 90}(t_{\rm lb})$, versus lookback time, $t_{\rm lb}$.
We measure these galaxy radii using all stars, or using only young stars (age $< 100 \Myr$), at a given time.
Each line and shaded region shows the mean and standard deviation across 11 FIRE-2 galaxies.
Including all stars, the typical growth over the last $10 \Gyr$ of $R_\star^{\rm 50}$ was modest ($40\%$), while the growth of $R_\star^{\rm 90}$ was more substantial ($91\%$).
Considering only recently-formed stars, the growth was even more significant, being $163\%$ for $R_{\rm \star,young}^{\rm 50}$ and $179\%$ for $R_{\rm \star,young}^{\rm 90}$.
This radial growth was especially pronounced over the last $\approx 8 \Gyr$, roughly when a typical disk started to form in FIRE-2.
\textit{These FIRE-2 MW-mass galaxies experienced physically meaningful ``inside-out" radial growth, with stars forming across a larger radial range over time.}
}
\label{fig:radius_v_time}
\end{figure}

We first quantify inside-out growth by tracking the radial size of the galaxies over time.
Figure~\ref{fig:radius_v_time} shows the galaxy radius versus lookback time, via the metrics of $R_{\star}^{50}$ and $R_\star^{\rm 90}$, the cylindrical radii that enclose $50\%$ and $90\%$ of the total mass of stars within an initial radial aperture of $30 \kpc$ comoving \citep[following][see also their Figure~2]{Bellardini2022}.
We measure these radii using all stars and using only young stars (age $< 100 \Myr$).
Both $R_\star^{\rm 50}$ and $R_\star^{\rm 90}$ grew over time, as star formation occurred over a larger radial range.
Over the last $10 \Gyr$, the average $R_{\star}^{50}$ for all stars increased from $2.5 \kpc$ to $3.5 \kpc$ ($40\%$), and $R_\star^{\rm 90}$ for all stars increased from $5.8 \kpc$ to $11.1 \kpc$ ($91\%$).
As a natural consequence of this inside-out growth, the galaxy radius for young stars grew significantly more than when using all stars.
The average $R_{\star}^{50}$ for young stars increased from $3.0 \kpc$ to $7.9 \kpc$ ($163\%$), while $R_\star^{\rm 90}$ for young stars increased from $6.1 \kpc$ to $17.0 \kpc$ ($179\%$).
The growth rate increased $\approx 8 \Gyr$ ago, when a typical disk began to form in these FIRE-2 galaxies \citep{McCluskey2023}.

Table~\ref{tab:galaxy_properties} lists these radii for all galaxies individually today.
Romeo, which has the earliest-forming disk, is the largest, with $R_\star^{\rm 90}$ for young stars of $23.8 \kpc$.

\begin{figure*}
\centering
\includegraphics[scale = 0.582]{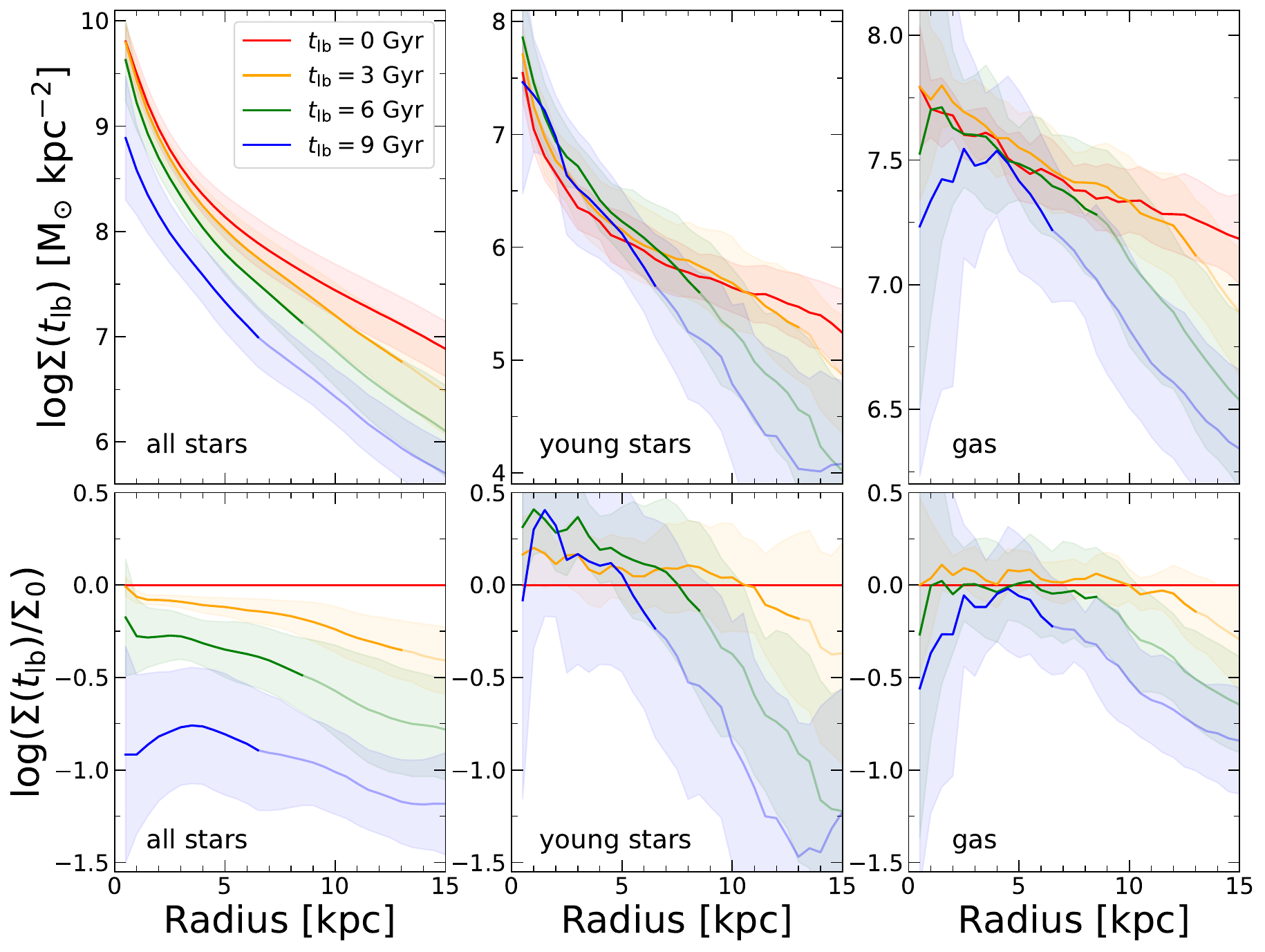}
\vspace{-5 mm}
\caption{
Our key metric of ``inside-out'' radial growth.
\textbf{Top}: Surface density, $\Sigma(R, t_{\rm lb})$, versus cylindrical radius in the disk, $R$, at various lookback times, $t_{\rm lb}$, for all stars (left), young stars (age $< 100 \Myr$, middle), and all gas (right).
Each line and shaded region shows the mean and standard deviation across 11 FIRE-2 galaxies.
Lines are lighter at $R > R_{\rm \star,young}^{\rm 90}(t_{\rm lb})$.
\textbf{Bottom}: Surface density at a given lookback time relative to today, at the same $R$, $\Sigma(R, t_{\rm lb}) / \Sigma(R, t_{\rm lb}=0)$.
All density profiles became less steep (shallower) towards today, especially at $t_{\rm lb} \lesssim 6 \Gyr$.
This reflects inside-out radial growth, with stars forming across a larger radial range as a galaxy evolves.
Because the \textit{realized} kpc-scale star-formation efficiency increases with total gas density, $\Sigma_{\rm SFR} \propto \Sigma_{\rm gas}^n$, with $n \approx 1.4$ in a Kennicutt-Schmidt-like relation, the shallowing of the profile towards today is even stronger for young stars than for gas.
\textit{Thus, young stars experience stronger inside-out radial growth than gas.}
}
\label{fig:density_profiles}
\end{figure*}

\begin{figure*}
\centering
\includegraphics[scale = 0.458]{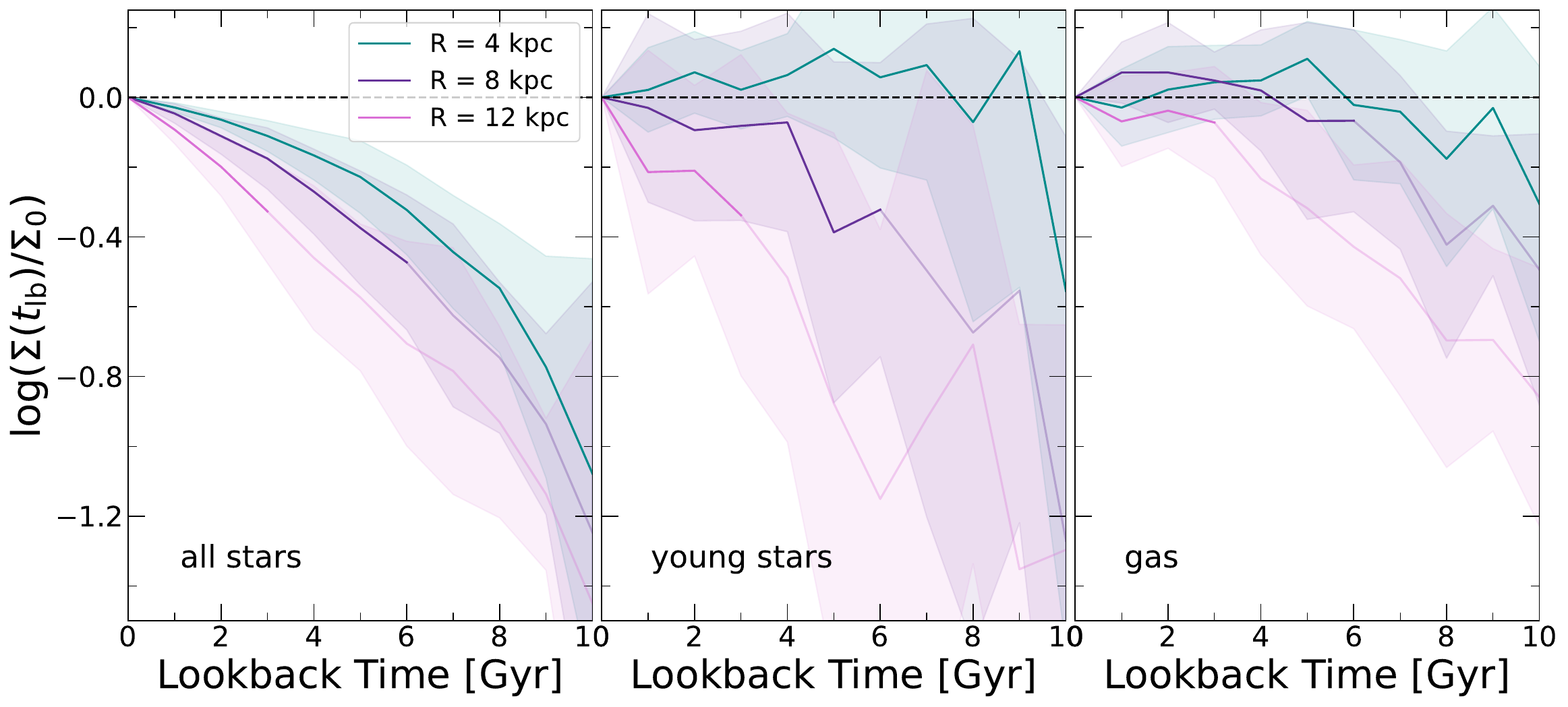}
\vspace{-5 mm}
\caption{
Similar to Figure~\ref{fig:density_profiles} (bottom), the fractional surface density at a given lookback time relative to today, $\Sigma(R, t_{\rm lb}) / \Sigma(R, t_{\rm lb}=0)$, but here we show it versus lookback time, $t_{\rm lb}$, at fixed physical radii, $R$, for all stars (left), young stars ($< 100 \Myr$, middle), and gas (right).
Each line and shaded region shows the mean and standard deviation across the 11 FIRE-2 galaxies.
Lines are lighter at $R > R_{\rm \star,young}^{\rm 90}(t_{\rm lb})$.
Considering all stars, $\Sigma(t_{\rm lb}) / \Sigma_{\rm 0}$ rises more quickly at larger $R$, especially over the last $\approx 6 \Gyr$.
\textit{This inside-out radial growth is stronger for gas, and given that the realized $\Sigma_{\rm SFR} \propto \Sigma_{\rm gas}^n$ with $n \approx 1.4$, it is even stronger yet for young stars.}
For example, at $R = 4 \kpc$, the surface densities of gas and young stars are relatively unchanged over the last $\approx 10 \Gyr$, while the surface densities at $R = 12 \kpc$ grew by a factor of $\approx 10$.
}
\label{fig:density_v_time}
\end{figure*}

A more detailed way of quantifying inside-out radial growth is via the evolution of the mass surface density profile.
Meaningful inside-out growth would correspond to the density increasing more rapidly, in a fractional sense, at larger $R$ than at smaller $R$.
In other words, inside-out growth means that the density profile became shallower over time.
This contrasts with self-similar growth, in which the fractional growth rate was the same at all $R$, and the steepness of the profile did not change.
We do not consider that to be inside-out growth.

Figure~\ref{fig:density_profiles} shows the mass surface density, $\Sigma(R)$, versus cylindrical radius in the disk, $R$, at various lookback times, $t_{\rm lb}$.
We show profiles for all stars (left), young stars (middle), and all gas (right).
The bottom panels show the ratio of the surface density at a given $t_{\rm lb}$ to the surface density today, at the same $R$.
For reference, we show $R > R_{\rm \star,young}^{\rm 90}(t_{\rm lb})$ as lighter lines.

Figure~\ref{fig:density_v_time} shows the same results as Figure~\ref{fig:density_profiles} (bottom), but instead versus lookback time, $t_{\rm lb}$ at fixed physical $R$.
Thus, we discuss both figures together.

For all components, the FIRE-2 galaxies on average show greater fractional growth at larger $R$ than at smaller $R$, at least over the last $\approx 6 \Gyr$, since the emergence of a (thin) disk in these galaxies.
The behavior at earlier times, in some cases before the onset of a disk, is more complicated, with some galaxies instead showing the most significant growth near $R \approx 0 \kpc$.
Appendix~\ref{sec:appendix} shows the history of each galaxy individually.

First, considering all stars (left panels), $\Sigma_\star(R)$ grew at all $R$ at all times.
At a given $R$, $\Sigma_{\star}$ grew more rapidly at earlier times, with fractional growth slowing as the galaxies evolved.
Before $\approx 6 \Gyr$ ago, the fractional growth was closer to self-similar at most $R$, but the last $\approx 6 \Gyr$ shows clear inside-out growth at all $R$.

Figure~\ref{fig:density_profiles} (middle) shows that inside-out growth occurred because significant star formation occurred \textit{over a larger range in $R$} as the galaxies evolved, and not because star formation halted at small $R$.
By contrast, at late times, the star formation surface density remained highest at small $R$, and the overall stellar density at small $R$ increased to today.
Inside-out growth occurred because of the more dramatic increase in star formation at larger $R$.

More relevant for the metallicity radial gradient is the evolution of young stars (middle panels) and total gas (right panels).
The evolution of gas drives the evolution of star formation, so the inside-out growth of stars emerges from inside-out growth of gas.
This arises primarily from cosmic gas accretion, and at late times this tends to occur more gradually and coherently into the outer disks of these FIRE-2 galaxies \citep{Trapp2022}.
Therefore, $\Sigma_{\rm gas}(R)$ grew significantly at large $R$, while it remained roughly constant at small $R$, as balanced by accretion, star formation, and regulation (outflows) from stellar feedback.

As a consequence, the star-formation surface density also increased at larger $R$, and the fractional growth of young stars is even more dramatic than total gas at large $R$.
As we discussed in Section~\ref{sec:background}, this is because the \textit{realized} kpc-scale star-formation rate density in FIRE increases superlinearly with the total gas density, $\Sigma_{\rm SFR} \propto \Sigma^{n}_{\rm gas}$, with $n \approx 1.4$, consistent with a Kennicutt-Schmidt-like relation \citep{Orr2018}.
Thus, the surface density profile of young stars became shallower more rapidly than that of the total gas.

Again, Figures~\ref{fig:density_profiles} and \ref{fig:density_v_time} show average trends, with galaxy-to-galaxy scatter, across our 11 FIRE-2 galaxies. Appendix~\ref{sec:appendix} shows these trends for each galaxy individually.
While inside-out radial growth occurs to varying degrees in almost all galaxies, a few exhibit growth that is nearly self-similar at all $R$.

\subsection{Evolution of radial gradients}
\label{sub:gradient_evolution}

\begin{figure}
\centering
\includegraphics[width = \columnwidth]{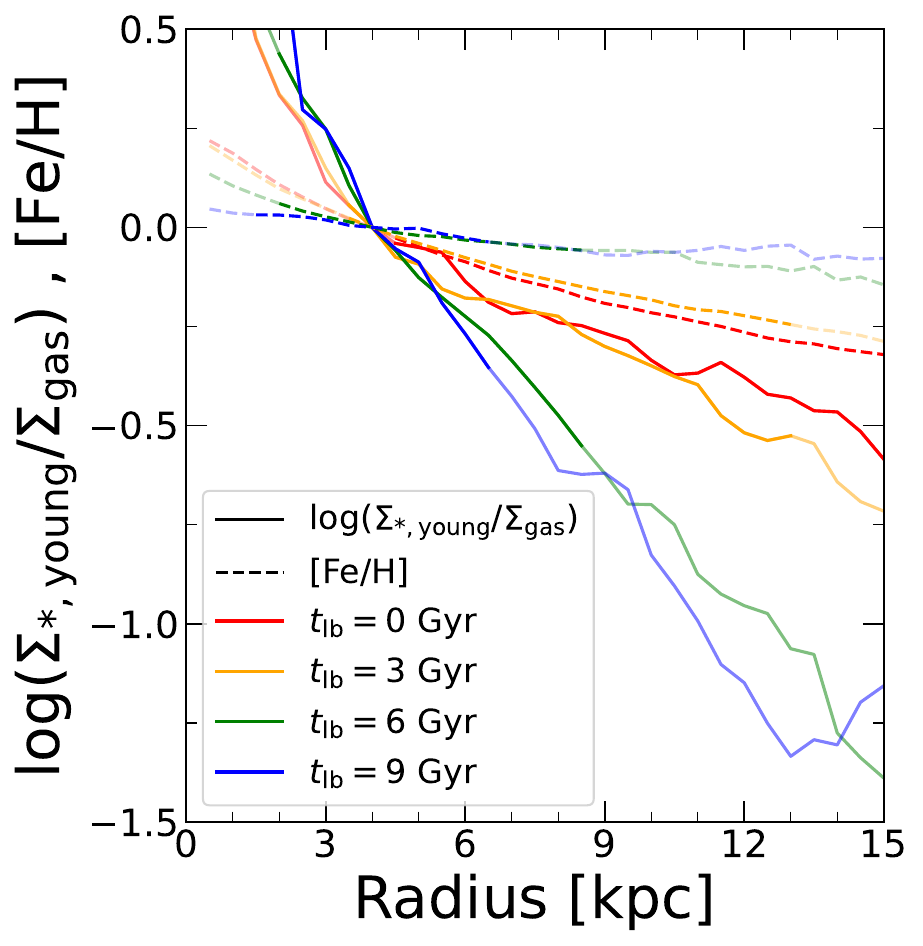}
\vspace{-5 mm}
\caption{
Two profiles versus cylindrical radius in the disk, $R$, at various lookback times, $t_{\rm lb}$.
First, dashed lines show [Fe/H]$(R)$ for young stars (age $< 100 \Myr$), which reflects the ISM at the same $t_{\rm lb}$.
Second, solid lines show the log ratio of the surface density of young stars to total gas, $\log( \Sigma_{\rm \star, young}(R) / \Sigma_{\rm gas}(R) )$.
To compare their shapes, we normalize all profiles to 0 at $R = 4 \kpc$.
Each line shows the mean across the 11 FIRE-2 galaxies.
If metals stayed where they were injected into the ISM from (young) stars, with no gas mixing/flows, then the ratio of young stars to gas would approximately determine the shape of [Fe/H]$(R)$ in the ISM at that time, and thus for stars at birth.
Both [Fe/H]$(R)$ and $\log( \Sigma_{\rm \star, young}(R) / \Sigma_{\rm gas}(R) )$ have similar profiles today (red), at least at $R \gtrsim 4 \kpc$.
\textit{However, the two profiles increasingly diverge at earlier times.
Thus, the ratio of the density of (young) stars to gas does not even qualitatively determine the evolution of the ISM metallicity radial profile.}
}
\label{fig:metallicity_density_profiles}
\end{figure}

\begin{figure}
\centering
\includegraphics[width = \columnwidth]{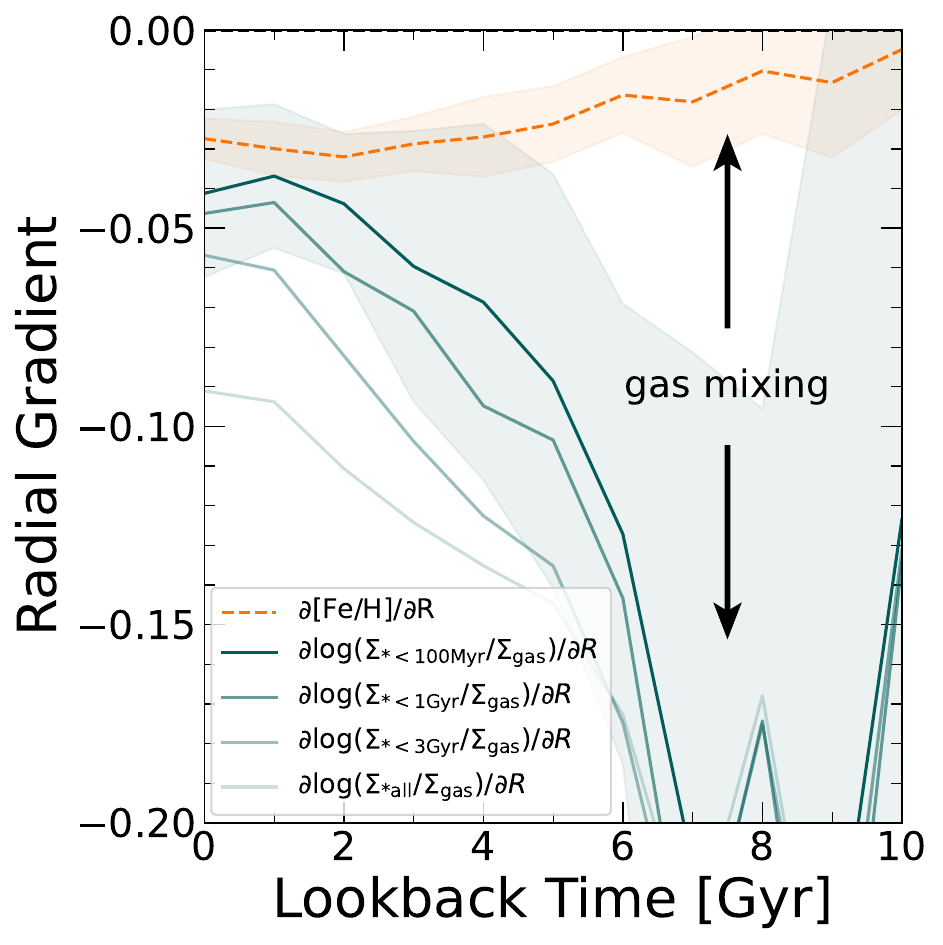}
\vspace{-5 mm}
\caption{
Evolution of radial gradients.
Here we measure all gradients across $R = (0.25 - 1) R^{\rm 90}_{\rm \star,young}(t_{\rm lb})$, at each lookback time, $t_{\rm lb}$.
Each line and each shaded region shows the mean and standard deviation across the 11 FIRE-2 galaxies.
Dashed orange line shows the metallicity gradient, via $\partial {\rm [Fe/H]} / \partial R$, of young stars (age $< 100 \Myr$) at each time, which reflects the gradient of the ISM at that time, and became steeper (more negative) towards today.
Green lines show $\partial \log( \Sigma_{\rm \star,young} / \Sigma_{\rm gas}) / \partial R$, which became flatter towards today, from the combination of inside-out radial growth and Kennicutt-Schmidt-like star-formation-efficiency as in Figures~\ref{fig:density_profiles} and \ref{fig:density_v_time}.
Including older stars leads to a steeper gradient in the density ratios, but all trends are qualitatively similar.
If metals stayed where they were injected into the ISM by (young) stars, then the metallicity gradient of the ISM, and of stars at birth, would track a green line.
\textit{While the two gradients are similar today, they increasingly diverge at earlier times.
Therefore, processes other than inside-out radial growth regulate the evolution of the metallicity radial gradient.
As we argue in Section~\ref{sub:mixing}, the key additional process is the evolution of metal mixing in gas.}
}
\label{fig:gradient_evolution}
\end{figure}

Having demonstrated that almost all of the FIRE-2 galaxies experienced inside-out radial growth, we reiterate expectations from Section~\ref{sec:background}, that if metals stayed where they were injected in the ISM, then Equation~\ref{eq:gradient} would determine the metallicity gradient.

However, one source of ambiguity is the appropriate choice of ``young'' regarding the ages of stars that are most relevant in determining \textit{spatial variations} in ISM metallicity across a galaxy, like the radial gradient.
In the FIRE-2 model, for a given stellar population, most metals are injected in the first $\approx 100 \Myr$.
This sets a lower limit to the relevant age.
Considering an extreme case, if all metals injected from stars older than this age have become uniformly mixed across the ISM, only these young stars would determine the metallicity gradient.
On the other extreme, if stars of all ages stayed at the same $R$ where they formed and all metals stayed where stars injected them, one should use the density profile of all stars, regardless of age.
Reality will lie somewhere between these extremes, depending on the formation history of a galaxy.
In general, metals that were injected many Gyr ago are more likely to have become mixed in the ISM (or ``lost'' into the CGM/IGM), so one would have to consider the time dependence of these processes and any characteristic timescales.
Therefore, below we explore our results using a range of age thresholds to determine ``young'' stars, and we show that our qualitative results are not sensitive to this choice.
Again, our fiducial definition of ``young'' stars is age $< 100 \Myr$.

As a reminder, in our analysis below, we measure $\Sigma_{\rm gas}(R)$ directly from each simulation, so at a given time it has already been shaped by processes like dilution from cosmic accretion, etc.
Furthermore, we measure $\Sigma_{\rm \star,young}(R)$ from the realized star-formation history from each simulation.
Given these inputs from each simulation, the profile $\log(\Sigma_{\rm \star,young}(R) / \Sigma_{\rm gas}(R))$ below thus provides a proxy for the strength of the metallicity profile, \textit{if} metals stay where they were injected from stars, as in Equation~\ref{eq:profile}.

Figure~\ref{fig:metallicity_density_profiles} shows two profiles versus cylindrical radius in the disk, $R$, at different lookback times, $t_{\rm lb}$.
First, dashed lines show [Fe/H]$(R)$ in young stars (ages $< 100 \Myr$), which closely resembles that of the star-forming ISM at that time, as we checked.
Second, solid lines show $\log(\Sigma_{\rm \star,young}(R) / \Sigma_{\rm gas}(R))$.
To compare their shapes, we normalize all profiles to 0 at $R = 4 \kpc$.
We show thicker lines at $R = (0.25 - 1)R^{\rm 90}_{\rm \star,young}(t_{\rm lb})$, which is the range over which we measure radial gradients, as we explain below.

If metals stayed at the same $R$ where stars injected them into the ISM, then the solid and dashed lines should track each other at a given $t_{\rm lb}$, according to Equation~\ref{eq:profile}.
Indeed, today they approximately do, as the red lines in Figure~\ref{fig:metallicity_density_profiles} show, consistent with a similar analysis in Figure~4 of \citet{Bellardini2022}.
Specifically, the red lines agree well at $R \gtrsim 4 \kpc$, though [Fe/H]$(R)$ is slightly shallower.
The two profiles increasingly diverge at $R \lesssim 4 \kpc$.
As we discuss in Section~\ref{sub:mixing}, these differences in inner slope likely arise from metal mixing in the ISM and its increase at small $R$, along with the saturation in enrichment at ${\rm [Fe/H]} \gtrsim 0.4$ dex, given the FIRE-2 yield model.
Related to this, \citet{Bellardini2022} found that the ``break'' $R$, beyond which the slopes of the profile becomes shallower, is similar for $\log( \Sigma_{\rm \star,young}(R) / \Sigma_{\rm gas}(R) )$ and [Fe/H]$(R)$ within a given FIRE-2 galaxy today.

The key result in Figure~\ref{fig:metallicity_density_profiles} is that the two profiles increasingly diverge at earlier times, when [Fe/H]$(R)$ was nearly flat (on average) and $\log( \Sigma_{\rm \star,young}(R) / \Sigma_{\rm gas}(R) )$ was even steeper.
\textit{The evolution of $\Sigma_{\rm \star,young}(R) / \Sigma_{\rm gas}(R)$ does not even qualitatively determine the evolution of [Fe/H]$(R)$ of the ISM in these FIRE-2 galaxies.}

The results of Section~\ref{sub:inside_out_growth} explain why the profile of $\Sigma_{\rm \star,young}(R) / \Sigma_{\rm gas}(R)$ became shallower over time.
$\Sigma_{\rm gas}(R)$ became shallower over time via inside-out radial growth largely from cosmic accretion, but $\Sigma_{\rm \star,young}(R)$ became shallower more quickly, given the superlinear (Kennicutt-Schmidt-like) realized star-formation efficiency in these FIRE-2 simulations.

By contrast, [Fe/H]$(R)$ became steeper over time, and its profile was nearly flat, on average, at $t_{\rm lb} \gtrsim 9 \Gyr$ ago.
Again, this trend in the ISM imprinted itself in the same trend versus $t_{\rm lb}$ for stars at birth, and this trend is broadly consistent with the trend for stars today versus their age as measured in the MW, as in Figure~\ref{fig:mw_compare}.

Figure~\ref{fig:gradient_evolution} shows the most important results of our analysis.
We directly compare the radial gradients of both [Fe/H]$(R)$ and $\log(\Sigma_{\rm \star,young}(R) / \Sigma_{\rm gas}(R))$ versus lookback time, $t_{\rm lb}$.
Unlike in Figure~\ref{fig:mw_compare}, here we measure the gradients across $R = (0.25 - 1)R^{\rm 90}_{\rm \star,young}(t_{\rm lb})$, excluding small $R$.
This is because, as Figure~\ref{fig:metallicity_density_profiles} showed, the profile of $\log( \Sigma_{\rm \star,young}(R) / \Sigma_{\rm gas}(R) )$ is not well fit by a single gradient, especially at late times, and because the profile of $\log( \Sigma_{\rm \star,young}(R) / \Sigma_{\rm gas}(R) )$ does not account for the possible saturation in enrichment at high metallicity, which can be relevant for the metal-rich inner galaxy.
Both profiles are better fit by a single gradient across $R = (0.25 - 1)R^{\rm 90}_{\rm \star,young}(t_{\rm lb})$.
That said, this choice does not qualitatively affect our results.
This radial range is also consistent with other analyses of radial gradients in the FIRE simulations (\citealt{Ma2017a, Sun2024}).

The orange dashed line in Figure~\ref{fig:gradient_evolution} shows the evolution of the metallicity gradient, $\partial {\rm[Fe/H]} / \partial R$, in young stars, which closely reflects that of the ISM.
Despite using a smaller radial range, these results are nearly identical to the orange line in Figure~\ref{fig:mw_compare}, given that the profile of [Fe/H]$(R)$ is nearly linear in $R$.
The solid green lines in Figure~\ref{fig:gradient_evolution} show the evolution of $\partial \log( \Sigma_{\rm \star,young} / \Sigma_{\rm gas} ) / \partial R$.
We show results using different age thresholds for ``young'' stars.
The darkest line shows our fiducial threshold of age $< 100 \Myr$, while the lighter lines show $< 1 \Gyr$, $< 3 \Gyr$, and all stars (no age threshold).
Given the inside-out radial growth of these galaxies, as in Figure~\ref{fig:density_profiles}, including increasingly older stars leads to a steeper profile of $\Sigma_{\rm \star}(R)$ and therefore a steeper gradient in $\partial \log(\Sigma_{\rm \star} / \Sigma_{\rm gas}) / \partial R$ \citep[see also Figure~4 in][]{Bellardini2022}.
This dependence on stellar age threshold becomes stronger over time, as galaxies contain older stars.
Nonetheless, for all age thresholds in determining ``young'' stars, the results are qualitatively consistent, that $\partial \log(\Sigma_{\rm \star,young} / \Sigma_{\rm gas}) / \partial R$ becomes shallower (flatter) over time.

Quantitatively, in Figure~\ref{fig:gradient_evolution}, the average $\partial {\rm[Fe/H]} / \partial R = -0.027$ dex kpc$^{-1}$ at $t_{\rm lb} = 0$, which shallows to be nearly flat, on average, at $t_{\rm lb} \gtrsim 9 \Gyr$.
By contrast, $\partial \log( \Sigma_{\rm \star,young} / \Sigma_{\rm gas} ) / \partial R$ ranges from $-0.041$ to $-0.091$ dex kpc$^{-1}$, depending on stellar age threshold, at $t_{\rm lb} = 0$, and steepens with increasing $t_{\rm lb}$, to $\approx -0.25$ dex kpc$^{-1}$ for all stellar age thresholds $\gtrsim 9 \Gyr$ ago.

The two gradients diverge at earlier times despite the expectations from the inside-out growth that these FIRE-2 galaxies experienced (Section~\ref{sub:inside_out_growth}).
\textit{Therefore, processes other than inside-out radial growth regulate the metallicity radial gradient.
As we argue in Section~\ref{sub:mixing}, the key additional process that explains the divergence between the two gradients and its dependence on time is the evolution of metal mixing in gas.
Specifically, strong mixing early on leads to nearly flat metallicity gradients, while relatively weak radial mixing today causes the two gradients to be similar.}

The MW shows evidence for an unusually early-forming disk \citep{BelokurovKravtsov2022, Conroy2022, XiangRix2022}, so one may wonder if these trends are meaningfully different for early-forming disks within our FIRE-2 sample.
We examined Romeo, our earliest-forming disk ($\approx 11 \Gyr$ ago, likely within $\approx 1 \Gyr$ of the MW), and its behavior in Figure~\ref{fig:gradient_evolution} sits within the standard deviation (shaded region) for both gradients at almost all $t_{\rm lb}$.
We therefore conclude that the results of Figure~\ref{fig:gradient_evolution} do not depend qualitatively on disk formation time, at least within FIRE-2.

In summary, FIRE-2 MW-mass galaxies have metallicity radial gradients in the ISM and thus in young stars that became steeper (more negative) over time.
This occurs despite nearly all of them forming radially inside-out, such that the naive expectation from Equation~\ref{eq:gradient} would be that their metallicity gradient became shallower (flatter) over time.

\subsection{Mixing of metals in the ISM}
\label{sub:mixing}

We now examine metrics of the (radial) mixing of metals in gas in these FIRE-2 simulations, including gas velocity dispersions, radial advection, and indications of radially-dependent galactic outflows, which would mediate the onset and evolution of metallicity radial gradients in the ISM.
We are interested in the effects of mixing on the \textit{azimuthally-averaged} radial gradient, which is sensitive not just to \textit{local} mixing via turbulence on small ($\lesssim 1 \kpc$ scales), but also to larger-scale mixing processes, such as azimuthally-dependent bulk flows and spiral arms \citep[for example,][]{Orr2023}.
Thus, we examine all metrics of gas mixing (such as the velocity dispersion) as \textit{azimuthally-averaged} quantities, around a $360^\circ$ annulus at fixed $R$.
We also examine these dynamics in the total gas, not just in cold star-forming gas, given that metals can be injected into and will mix across different phases of the ISM.
That said, we investigate trends in both the total gas and in metal-rich gas, to explore potential differences.

\begin{figure*}
\centering
\includegraphics[scale = 0.518]{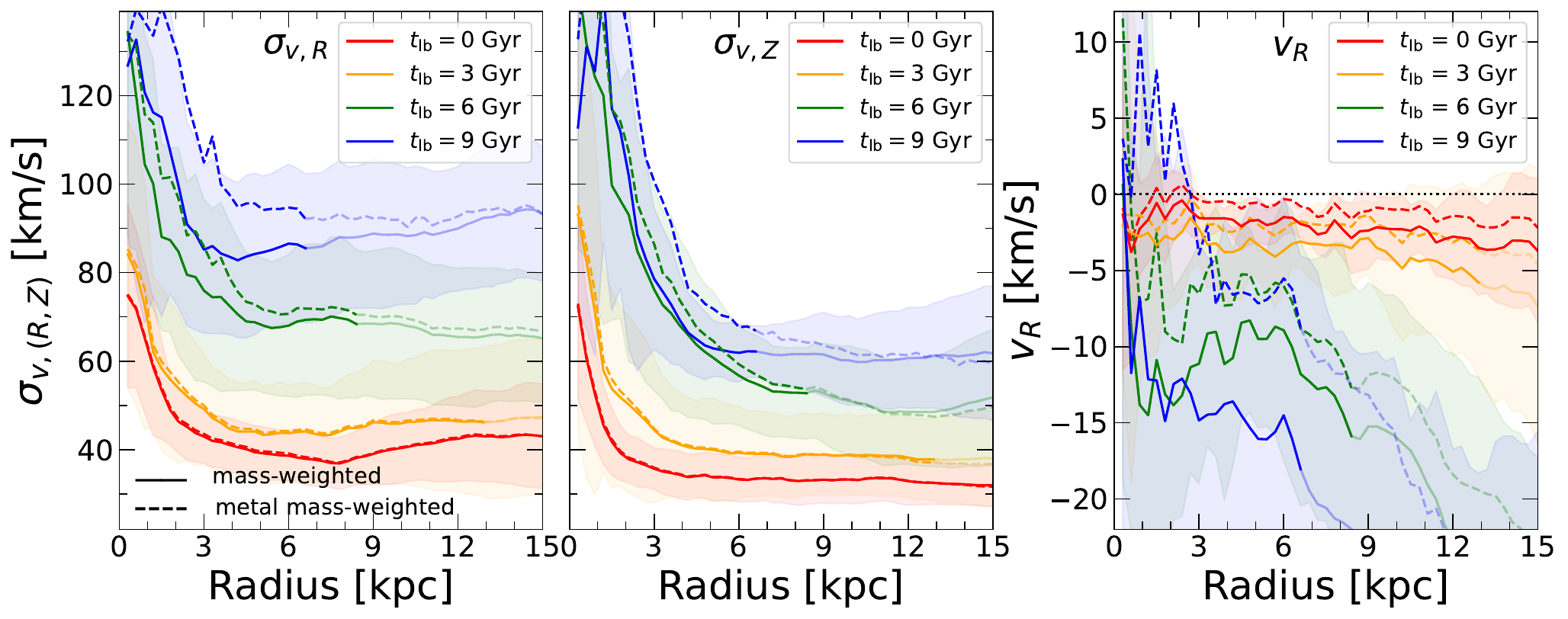}
\vspace{-5 mm}
\caption{
Metrics for metal mixing in the ISM, versus cylindrical radius in the disk, $R$, at various lookback times, $t_{\rm lb}$.
Because we are interested in the effect on the radial gradient, we compute all quantities across a $360^\circ$ annulus, not just in a local patch of the ISM.
Each line and shaded region shows the mean and standard deviation across 11 FIRE-2 galaxies.
Solid (dashed) lines show velocities weighted by the total (Fe) mass of each gas cell.
\textbf{Left}: The radial velocity dispersion, $\sigma_{\rm v,R}$, was significantly higher, $2 - 3 \times$, in the past.
This reflects higher gas turbulence from higher rates of both accretion/mergers and star formation/feedback, which led to stronger radial mixing of metals at earlier times.
The dispersion is slightly higher in metal-rich gas, likely because the injection of metals corresponds to the injection of turbulence from feedback.
This contributed to mixing away any strong metallicity radial gradient that would otherwise form (Figure~\ref{fig:gradient_evolution}).
\textbf{Middle}: Same, but for the vertical velocity dispersion, $\sigma_{\rm v,Z}$, which shows the same trends as for $\sigma_{\rm v,R}$.
This dynamical evolution drives the ``upside-down'' vertical growth of these galaxies.
Furthermore, the strong increase in $\sigma_{\rm v,Z}$ at smaller $R$ suggests radially-dependent galactic wind velocities that could eject metals preferentially from the inner galaxy.
\textbf{Right}: Mean radial velocity, $v_{\rm R}$, as a metric of net advection of gas.
Gas flowed radially inward more strongly at earlier times, when accretion and gas consumption (star formation) rates were higher, but the gas became more static, with $v_{\rm R} \approx 0 \kms$, over the last $\approx 3 \Gyr$.
Furthermore, at earlier times, the galaxies experienced significant offset in advection between metal-rich and overall gas.
\textit{The (radial) mixing of metals in gas---through a combination of turbulence, azimuthally-dependent flows, net radial flows, and galactic winds---declined as each galaxy evolved, its disk settled, and it experienced upside-down growth.
We conclude that this mixing is the primary reason why the metallicity radial gradient became steeper (more negative) over time, despite the inside-out radial growth of these galaxies.}
}
\label{fig:velocity_dispersion}
\end{figure*}

First, we investigate the radial and vertical velocity dispersions.
These serve as metrics of the overall mixing of gas on galactic scales, relevant for determining the metallicity radial gradient.
Again, we measure these as azimuthally-averaged (not local) velocity dispersions.
Therefore, many factors can influence this velocity dispersion in gas, including: turbulence; perturbations from giant molecular clouds, spiral arms, and bars; stellar feedback and galactic winds; accretion, mergers, and interactions with satellite galaxies.
We also average these measurements for each galaxy across $10$ consecutive snapshots centered at $t_{\rm lb}$, to ensure the velocities and dispersions are representative across time.

Figure~\ref{fig:velocity_dispersion} (left) shows the gas radial velocity dispersion, $\sigma_{\rm v,R}$, versus cylindrical radius, $R$, in bins of lookback time, $t_{\rm lb}$.
Solid lines show all gas, weighting each cell by its total mass, and dashed lines show metal-rich gas, weighting each cell by its mass in Fe.
$\sigma_{\rm v,R}$ is systematically slightly higher for metal-rich gas, likely because stellar feedback both sources the metals and drives additional turbulence in gas.

At all times, $\sigma_{\rm v,R}$ increases at smaller $R$, especially $R \lesssim 3 \kpc$.
This increase helps to explain why in Figure~\ref{fig:metallicity_density_profiles} the profile of [Fe/H]$(R)$ is much shallower than $\log(\Sigma_{\rm \star,young}(R) / \Sigma_{\rm gas}(R) )$ at $R \lesssim 3-4 \kpc$: the enhanced radial mixing in the inner galaxy leads to a shallower realized metallicity gradient.
That said, as \citet{Bellardini2022} explored, the profile of [Fe/H]$(R)$ does show a modest upturn at a similar $R$ that corresponds to the upturn in $\log( \Sigma_{\rm \star,young}(R) / \Sigma_{\rm gas}(R) )$.

Most importantly, Figure~\ref{fig:velocity_dispersion} (left) shows that $\sigma_{\rm v,R}$ was systematically higher in the past by a factor of $2 - 3$, from $\approx 40 - 50$ km s$^{-1}$ today to $\approx 90 - 110$ km s$^{-1}$ at $t_{\rm lb} = 9 \Gyr$.
This is consistent with previous analyses of ISM disk formation/settling in these FIRE-2 simulations, and most of these galaxies were not dominated by rotational motion until $\approx 8 \Gyr$ ago \citep[for example,][]{Gurvich2023, McCluskey2023, Yu2023}.
Again, we examine the \textit{azimuthally-averaged} ($360^\circ$ around an annulus) velocity dispersion of the \textit{total} gas, not just the local dispersion in the cold, molecular, and/or star-forming gas, as some previous works did, so the velocity dispersions in Figure~\ref{fig:velocity_dispersion} are significantly larger \citep[see discussion in][]{McCluskey2023}.
This high radial velocity dispersion at early times efficiently mixed the metals in the ISM and prevented a strong metallicity radial gradient from forming.
The gas velocity dispersion declined as the disk settled and $v_{\rm \phi} / \sigma_{\rm v}$ increased, leading to weak radial mixing and allowing a metallicity gradient to form that approximately tracks $\log(\Sigma_{\rm \star,young}(R) / \Sigma_{\rm gas}(R))$, as in Figures~\ref{fig:metallicity_density_profiles} and \ref{fig:gradient_evolution}.

Another contributor to gas mixing is the net radial advection of (metal-rich) gas, which can influence the metallicity radial gradient in various ways \citep[for example,][]{Sharda2021}.
As a metric of this, Figure~\ref{fig:velocity_dispersion} (right) shows the mean radial velocity, $v_{\rm R}$, for both total gas and metal-rich gas.
At late times, $t_{\rm lb} \lesssim 3 \Gyr$, the net radial advection is weak, being inward at a few km s$^{-1}$.
It is slightly stronger in total gas than in metal-rich gas, consistent with influx of pristine gas from cosmic accretion \citep{Trapp2022}.
However, as with the velocity dispersion, at earlier times $v_{\rm R}$ was more significant overall, and the advection was stronger at larger $R$, consistent with higher rates of cosmic accretion at earlier times.
Again, we see stronger inflow rates in total gas than in metal-rich gas, and at $t_{\rm lb} \gtrsim 9 \Gyr$, the metal-rich gas has positive $v_{\rm R}$ at small $R$, showing signs of radial outflows, even though the overall gas is net inflowing there.
The strong reduction in gas radial advection over time and the significant offset between overall and metal-rich gas at early times points to additional ``mixing'' that could contribute to washing out the formation of a strong gradient at early times.
That said, the net radial velocity in Figure~\ref{fig:velocity_dispersion} (right) is an order of magnitude smaller at a given $t_{\rm lb}$ than the radial velocity dispersion (left), which suggests that the dispersion is more important than net advection in regulating the metallicity gradient.
We defer a more in-depth study on the relative roles of advection and dispersion to future work.

Another effect that could modify the radial distribution of metals in the ISM is galactic winds.
Previous work has shown that these FIRE-2 simulations launch strong winds, particularly at early times \citep{Pandya2021}, which can carry significant metals with them.
This would affect the metallicity radial gradient if outflows launched metals in a radially-dependent way, and/or if the reaccretion of metal-rich winds back into the ISM was radially dependent \citep[see for example][]{Orr2022a, Orr2022b}.
This is another form of ``mixing'', though in this case mixing metals between the ISM and CGM.
While many works examined the overall metal outflow rates, few works have examined their radial dependence.
A detailed investigation is beyond the scope of this work, but as a basic metric, Figure~\ref{fig:velocity_dispersion} (middle) shows the vertical velocity dispersion of gas, $\sigma_{\rm v,Z}$.
It shows generally the same trends as $\sigma_{\rm v,R}$, including a strong increase with $t_{\rm lb}$ and a higher dispersion in metal-rich gas.
However, $\sigma_{\rm v,Z}$ shows an even stronger increase towards smaller $R$.
This might suggest that stellar feedback drives preferentially more metals out of the inner galaxy via galactic winds.
If these outflows and/or if their reaccretion varies with $R$, then this would lead to a radially-dependent metal loss that would flatten the metallicity gradient, especially at earlier times.

\subsection{Metal mixing versus inside-out growth}

In summary, our analysis of gas dynamics, including the radial and vertical velocity dispersion and net radial advection, indicates that the mixing of metals in the ISM of these MW-mass galaxies declined as they evolved vertically ``upside-down'' and formed thinner disks with more ordered rotation.
We argue that this evolution of metal mixing is the primary reason why the ISM metallicity radial gradient in these FIRE-2 galaxies became steeper over time, in contrast with the expectation from Equation~\ref{eq:gradient} given that these galaxies generally grew radially inside-out.
Specifically, metal mixing dominates over inside-out growth in determining the evolution of the metallicity gradient.
The strong mixing at early times led to a flat metallicity gradient, on average, despite the galaxies otherwise ``wanting'' to form a strong gradient according to Equation~\ref{eq:gradient}.
By contrast, the relatively weak mixing today means that Equation~\ref{eq:gradient} approximately sets the metallicity gradient today, on average, as in Figures~\ref{fig:metallicity_density_profiles} and \ref{fig:gradient_evolution}.
Given that these 11 FIRE-2 galaxies show the same qualitative trends as the MW, that younger stellar populations exhibit steeper (more negative) metallicity radial gradients, we argue that the scenario above reflects the ISM history of the MW and of typical MW-mass disk galaxies today.

Our results agree with the analysis of the FIRE-1 simulations in \citet{Ma2017a}.
They analyzed a much larger range in galaxy masses across $z = 0 - 2$ and found that the steepness of the ISM metallicity radial gradient correlates with (increases) most strongly with the degree of rotational support, $v_{\rm \phi} / \sigma_{\rm v}$, of the disk, a trend that they showed agrees with observational results.
They also interpreted this as arising from the strength of metal mixing in gas.
Because $v_{\rm \phi} / \sigma_{\rm v}$ increases throughout the histories of the MW-mass galaxies we analyze \citep{McCluskey2023}, our results reflect the same physical drivers.
Indeed, \citet{Bellardini2022} showed a significant correlation for young stars in these FIRE-2 galaxies between the strength of metallicity radial gradients and $v_{\rm \phi} / \sigma_{\rm v}$ today.

Our results also agree with recent JWST observations of 26 star-forming galaxies at $z \sim 1$ with masses comparable to the progenitors of MW-mass galaxies today \citep{Ju2024}.
They showed that, at these early times, the galaxies generally have flat radial gradients in ISM metallicity, and furthermore, that the gradients tend to be steeper (more negative) in more massive, ``diskier'' galaxies.
They also showed that these same FIRE-2 galaxies agree well with these observations at $z \sim 1$.

In future work, we will examine quantitatively, using analytic models applied to our simulation results, how the trends of metal mixing in Figure~\ref{fig:velocity_dispersion} can reconcile the discrepant evolution in 
Figure~\ref{fig:gradient_evolution} of $\partial {\rm [Fe/H]} / \partial R$ against the expectations from $\partial \log( \Sigma_{\rm \star,young} / \Sigma_{\rm gas}) / \partial R$ as in Equations~\ref{eq:profile} and \ref{eq:gradient}.

\section{Summary and Discussion}
\label{sec:Summary_and_Discussion}

\subsection{Summary}
\label{sub:summary}

We analyzed 11 MW-mass galaxies from the FIRE-2 suite of cosmological simulations to understand the evolution of the metallicity radial gradient in the ISM and in stars at birth.
These FIRE-2 simulations show a similar trend today as observed in the MW, that older stellar populations show systematically shallower (flatter) metallicity radial gradients. 
In FIRE-2, this trend emerges primarily because the metallicity gradient in the ISM became steeper (more negative) as the galaxies evolved.
In particular, in FIRE-2 this trend does \textit{not} arise from the radial redistribution of stars after birth, though radial redistribution does cause the gradient to be somewhat shallower than at birth.
Previous works argued that the ``inside-out" radial growth of a galaxy should regulate the evolution of its metallicity radial gradient, specifically, causing the ISM gradient to become shallower over time.
Therefore, we quantified the radial growth of gas, young stars, and all stars in these FIRE-2 galaxies.
We also examined metrics for the mixing of metals in the ISM and how they vary with radius and over time, which also lead to the ``upside-down'' vertical growth of these galaxies.
Ultimately, we sought to understand how the combination of inside-out and upside-down growth, including metal mixing in a self-consistent cosmological framework, regulate the evolution of metallicity radial gradients throughout the histories of MW-mass galaxies.

Our main results are as follows.

\begin{itemize}

\item All but one of these 11 galaxies experienced physically meaningful inside-out radial growth, in gas and in stars.
A typical galaxy started significant inside-out radial growth $\approx 6 \Gyr$ ago, though with significant galaxy-to-galaxy scatter.
Over the last $\approx 10 \Gyr$, $R_\star^{\rm 90}$ of all (young) stars increased on average by $\approx 90\%$ (180\%).

\item This inside-out radial growth arises because the surface density profile of gas became shallower over time, given the nature of gas accretion.
Furthermore, because the \textit{realized} kpc-scale relation between star formation and total gas in the FIRE simulations follows a Kennicutt-Schmidt-like scaling, $\Sigma_{\rm SFR} \propto \Sigma_{\rm gas}^n$, with $n \approx 1.4$ \citep{Orr2018}, the surface density profile of star formation (and thus young stars) is both steeper than and evolved more rapidly than that of gas.
As a result, the profile of the ratio of young stars to gas, $\Sigma_{\rm \star,young}(R) / \Sigma_{\rm gas}(R)$, became shallower (flatter) over time.

\item If metals stayed where they were injected into the ISM from (young) stars, then the gradient of $\log( \Sigma_{\rm \star,young} / \Sigma_{\rm gas} )$ largely would determine the metallicity radial gradient of the ISM, and thus of stars at birth, as in Equation~\ref{eq:gradient}, as some works have argued.
This relation approximately holds true today, when these galactic disks are settled and metal mixing in gas is modest.
However, $\partial \log(\Sigma_{\rm \star,young} / \Sigma_{\rm gas}) / \partial R$ became shallower (flatter) over time from inside-out radial growth, while $\partial {\rm[Fe/H]} / \partial R$ became steeper over time.

\item We argue that the diverging evolution of these gradients arises primarily from the evolution of the mixing of metals in gas in these FIRE-2 galaxies, likely from declining rates of both gas accretion/mergers and star formation/feedback, which also lead to ``upside-down'' vertical growth of the disks.
We show a strong decline over time in the radial and vertical velocity dispersion of (metal-rich) gas, which may also indicate radially-dependent ejection of metals via galactic winds.
We also demonstrate a strong decline over time in the net radial advection of gas, including declining offsets between metal-rich and overall gas.
We conclude that strong metal mixing at early times washed out any metallicity gradient in the ISM, and that the strong decline in ISM metal mixing over time as the disk settled is the primary reason why the ISM metallicity radial gradient became stronger (steeper) over time in FIRE-2.

\item Given that these 11 FIRE-2 galaxies show the same general trend as the MW, that younger stellar populations exhibit steeper (more negative) metallicity radial gradients, we argue that the scenario above reflects the ISM history of the MW and of typical MW-mass galaxies today.

\end{itemize}

\subsection{Caveats}
\label{sub:caveats}

We first discuss caveats to our results.
We analyzed a sample of $11$ MW-mass galaxies with a simple set of selection criteria: cosmologically isolated or in a LG-like pair, with $M_{\rm 200 m} \approx 1 - 2 \times 10^{12} \Msun$.
Thus, these galaxies sample random formation histories within those criteria.
As such, they are not perfect analogs of the MW.
\citet{Graf2024} found, for example, that these FIRE-2 galaxies generally have shallower metallicity radial gradients, by $\approx 0.015 - 0.04$ dex $\kpc^{-1}$, than the MW, though, as \citet{Bellardini2021} found, they agree with measurements of M31, and are comparable to or steeper than many measurements of the ISM in nearby MW-mass galaxies.
These comparisons suggest that the MW has an unusually steep metallicity radial gradient, at least today.
One possible contributor could be distance uncertainties in measurements of the MW's ISM: \citet{Donor2018} found that the distance catalog used can affect observed radial gradients by as much as $40\%$.
Another possibility, as we previously discussed, is that the MW's disk settled unusually early, and as a result, it has an unsusually dynamically thin/cold disk today \citep{McCluskey2023} with an unusually steep metallicity gradient.
Using this FIRE-2 sample, \citet{Bellardini2022} did find a weak correlation between the strength of radial gradients and disk settling time.

Furthermore, these FIRE-2 simulations do not model potentially important physical processes.
They do not include self-consistent cosmic rays injection and transport \citep[see discussions in][]{Hopkins2018, Wetzel2023}.
As \citet{Hopkins2020} showed, cosmic rays can provide a source of non-thermal pressure in the CGM that prevents gas accretion, which could affect both the amount of inside-out growth and the degree of turbulence/mixing in the ISM.
That said, analyses of FIRE-2 simulations that included cosmic rays show that they do not significantly alter disk dynamics \textit{at fixed stellar mass} \citep[for example,][]{Su2017, McCluskey2023}.
Moreover, these FIRE-2 simulations do not include supermassive black holes and AGN feedback.
AGN could be significant, especially in regulating the gas supply in the inner galaxy, which in turn could reduce star formation there, though they are less likely to affect dynamics and star formation near the Solar annuli of MW-mass galaxies \citep{MercedesFeliz2023, Wellons2023}.
Thus, by reducing star formation in the inner galaxy at late times, AGN could lead to even stronger inside-out radial growth.
Furthermore, some simulations \citep[for example,][]{Irodotou2022} indicate that AGN feedback could affect the galaxy-wide dynamics and structural properties of MW-mass galaxies.

Additionally, our results are based on the set of stellar nucleosynthesis models in FIRE-2, which are themselves uncertain \citep[see][]{Muley2021, Gandhi2022, Hopkins2023}.
However, as previous work showed \citep{Bellardini2021, Bellardini2022, Graf2024}, the trends in metallicity radial gradients that we examine hold true across all metals tracked in FIRE-2 (including C, N, O, Mg), despite their contributions from different channels of stellar evolution.
Thus, we consider it unlikely that uncertainties in stellar nucleosynthesis would qualitatively change our results.

Finally, we only focus on the evolution of radial gradients over long timescales and averaged over our sample of $11$ galaxies.
We do not examine fluctuations in gradients for individual galaxies over short timescales.
In particular, \citet{Ma2017a} and \citet{Sun2024} found that the metallicity gradients in FIRE galaxies at high redshifts can vary on short timescales, including brief periods with positive radial gradients, broadly consistent with the observations in \citet{Ju2024}.

\subsection{Inside-out radial growth}
\label{sub:inside_out_compare}

Our results on inside-out radial growth agree with nearly all investigations of cosmological simulations of galaxies at these masses.

\citet{Brook2012} analyzed a cosmological zoom-in simulation of a lower-mass galaxy ($\Mstar = 8 \times 10^{9} \Msun$), finding that the mass growth rate at large radii increased for younger stars, via inside-out growth.
\citet{Bird2012} analyzed the Eris simulation, and \citet{Bird2021} analyzed a GASOLINE simulation (h277), both cosmological simulations of MW-mass galaxies, finding that the radius of the star-forming disk increased with time.
Similarly, \citet{Agertz2021} analyzed the VINTERGATAN cosmological simulation of a single MW-mass galaxy and found the half-mass radius of its stars to increase with time.
They also showed that the surface density profile for stars shallows with time.
Analyzing $5$ galaxies from the NIHAO-UHD cosmological simulations, \citet{Buck2020} found that younger stars have larger scale radii in all but one of their galaxies.
Furthermore, analyzing one FIRE-2 galaxy (m12i), \citet{Carillo2023} found that the birth radius for older stars peaked at smaller R compared to younger stars, and also that the width of the distribution increased for younger stars, both of which reflect inside-out growth.
Some works also looked at the ages of stars versus radius.
Using the Auriga simulations, \citet{Iza2024} found stellar age to generally decrease with radius in MW-mass galaxies, consistent with inside-out growth, though some of their galaxies did not show this trend.

While these theoretical works agree that galaxies generally grow inside-out, observational evidence in external galaxies is more mixed.
Using the Hubble Frontier Fields, \citet{Tan2024} found the stellar surface density profile to grow in lockstep in MW analogs, suggesting self-similar, and not inside-out, growth.
Similarly, \citet{Hasheminia2022} found the stellar half-mass radius of MW-mass progenitor galaxies to remain roughly constant over the past $10 \Gyr$, despite stellar mass increasing by $\approx 1$ dex, and also that the stellar surface density profile grew in lockstep at all radii with time, both of which imply no inside-out growth.
On the other hand, \citet{Buitrago2024} measured $1040$ massive ($M_\star > 10^{10} \Msun$) disk galaxies and found that the size of the galaxies increased by a factor of $2$, thereby showing inside-out growth, and disagreeing with \citet{Hasheminia2022} based on the way one measured galaxy ``size".
Using the MaNGA survey, \citet{IbarraMedel2016} found that at late evolutionary stages, most ($\sim 70 \%$) galaxies showed inside-out growth in stars. Also using MaNGA, \citet{Pan2023} found stellar age to decrease with radius, as did \citet{Mun2024} using the MAGPI survey, a trend consistent with inside-out radial growth.
Recently, using the 3D-HST survey, \citet{Hasheminia2024} examined the mass dependence of inside-out radial growth, finding that only galaxies with $\Mstar \gtrsim 10^{10.7} \Msun$ today (comparable to the MW) show evidence for inside-out growth.
At that mass, they found a growth in $R_{80}$ (of all stars) of $28 - 44\%$ over the last $8 \Gyr$.

Many works also measured inside-out radial growth of the MW across its stellar populations.
\citet{Frankel2019} used a model in conjunction with APOGEE and found the half-mass radius of stars in the MW increased over time, evidence of inside-out growth.
Furthermore, \citet{Wang2024} analyzed $59,237$ stars in the GALAH DR3 survey and found that the birth radius distribution for old stars peaks at smaller $R$ compared to younger stars, and also that the width of the distribution increases for younger stars.
Similarly, \citet{Anders2023} analyzed 3060 red giant stars from APOGEE DR14 and found that older stars in the MW tend to reside at smaller radii.
Using APOGEE DR17, \citet{Lian2024} used a method for determining Galaxy size that uses the non-parametric surface brightness profile of the disk, and they found that the cumulative half-mass radius of the MW increases with decreasing stellar age.
They also found that $\Sigma_{\star}(R,t_{\rm lb})$ shallowed with time as mass growth increased at larger radii, indicating inside-out growth in the MW.

\subsection{Evolution of metallicity radial gradients}
\label{sub:shallow_vs_steep_compare}

In Section~\ref{sec:results} we showed that metallicity radial gradients in our $11$ MW-mass galaxies became steeper over time.
However, some theoretical models and simulations show the opposite trend, and while the literature shows consensus that simulated galaxies generally grow inside-out, there is no consensus regarding how metallicity radial gradients evolve.

Among works that agree with our results is \citet{VK2020}, which analyzed a cosmological zoom-in simulation of a single MW-mass galaxy and found a steepening stellar [Fe/H] radial gradient at later times.
Furthermore, \citet{Chiappini2001} developed an elemental evolution model that showed metallicity radial gradients steepen over time.
They argued that this occurs from infalling metal-poor gas in the outer galaxy at later times, which differs from our interpretation.
Similarly, \citet{Sharda2021} developed an analytical framework that argued for a steepening metallicity radial gradient over time.
Additionally, \citet{Tissera2022} used the EAGLE simulation to analyze $957$ galaxies with $M_\star > 10^9 \Msun$ and found gas-phase [O/H] radial gradients to have essentially no evolution ($\approx 0.001$ dex kpc$^{-1} / \delta z$) with redshift.
However, that work tracked galaxies of the same mass over cosmic time, as opposed to studying the histories of individual galaxies as we did, which complicates the comparison, given expectations (at least for FIRE) that the strength of radial gradients correlates with galaxy mass at a given redshift \citep[for example,][]{Ma2017a}.

Observationally, \citet{Cheng2024} analyzed $238$ star-forming galaxies at $z = 0.6 - 2.6$ to study [O/H] gas-phase radial gradients, and they found that galaxies have flat, and sometimes \textit{positive} radial gradients at early times, which tend to steepen in a negative direction over time, in agreement with our results.
Furthermore, \citet{Ju2024} studied radial gradients in ISM metallicity in $26$ galaxies at $z = 0.5 - 1.7$ using JWST and found typically nearly flat (and sometimes positive) radial gradients at these redshifts, with significantly negative gradients occurring mostly in the more massive, ``diskier'' galaxies in their sample.
They compared directly with these same FIRE-2 simulations, showing good agreement with their observed galaxies at $z \sim 1$.

Some theoretical works show a diversity of evolutionary paths for metallicity gradients.
\citet{Lu2022b}, for example, studied [Fe/H] radial gradients in the ISM using $4$ simulations from the NIHAO-UHD suite and found only mild evolution over time when measured across $R = 0 - 10 \kpc$, with some galaxies showing mild shallowing and others showing mild steepening over time. Furthermore, the VINTERGATAN cosmological simulation of a single MW-mass galaxy shows an [Fe/H] gradient which starts strong and shallows with time in its ISM \citep{Agertz2021}, while the [Fe/H] gradient in young stars begins flat and strengthens with time \citep{Renaud2024}.

Among works using simulations that disagree with our results is \citet{VK2018}, who analyzed cosmological simulations of $10$ MW-mass galaxies, specifically, [O/H] profiles in the ISM, and found gradients measured across $R \approx 0 - 15 \kpc$ to be strongest at early times, which shallow with time.
Using $4$ galaxies from the NIHAO-UHD suite of cosmological simulations, \citet{Buck2023} also found steeper metallicity radial gradients in gas at earlier times, seemingly in tension with \citet{Lu2022b}.
However, they measured the gas-phase gradient across a fixed radial range across cosmic time of $R = 2.5 - 17.5 \kpc$, which differs significantly from our measuring such gradients only in the star-forming ISM, that is, out to radii where significant stars form, $\approx R_{\rm \star,young}^{\rm 90}$, which may contribute to these different trends.
Furthermore, \citet{Acharyya2024} found gas-phase metallicity gradients to generally begin steep and shallow with time in the FOGGIE simulations.
Using the larger-volume TNG50 simulation, \citet{Hemler2021} studied $1595$ galaxies with $M_\star = 10^{9 - 11} \Msun$, and found gas-phase [O/H] radial gradients to be steeper in galaxies at higher redshift.
As with \citet{Tissera2022}, however, that work tracked galaxies of the same stellar mass over cosmic time, which again complicates the comparison with our work.


\citet{Minchev2018} developed an analytical method for determining stellar birth radii in the MW by extrapolating the evolution of the ISM metallicity with radius and time.
They argued that as a consequence of inside-out growth and the resultant increasing fractional stellar mass at larger radii with time, metallicity radial gradients began strong and shallowed with time, but appear today to be flat in old stellar populations, as many observational works find, because of radial redistribution over time.
While we agree that radial redistribution acts to flatten radial gradients in mono-age stellar populations, we disagree with the expectation of \citet{Minchev2018} that the radial gradient in the ISM was steeper in the past.

As we previously discussed, mixing via turbulence in metal-rich gas plays a key role in metallicity gradient evolution in FIRE-2.
Almost all cosmological simulations agree that turbulence was higher in the past.
\citet{McCluskey2023}, for example, found that in FIRE-2, the radial velocity dispersion, $\sigma_{\rm v,R}$, of stars at their formation increases with increasing stellar age for ages $\lesssim 9 \Gyr$, which agrees with our Figure~\ref{fig:velocity_dispersion} (left).
Furthermore, using the Horizon-AGN and NewHorizon hydrodynamical simulations, \citet{Han2024} found $\sigma_{\rm v,R}$ to be systematically higher for old stars than for young stars, also in agreement with our findings.

Although simulations agree the ISM turbulence generally declines over time, they disagree about whether the evolution is strong enough to counteract inside-out radial growth as the dominant effect for the evolution of metallicity gradients.
\citet{Pilkington2012} and \citet{Gibson2013} found that the ``weaker" stellar feedback model of the MUGS simulations tends to produce strong [O/H] radial gradients in gas at early times that then flatten over time, and that the ``enhanced" feedback model of the MaGICC simulations (otherwise identical to MUGS) tends to produce weak [O/H] radial gradients in gas at early times, that then steepen over time.
These results agree with ours, that the evolution of mixing/turbulence in gas is a key determinant in the evolution of metallicity gradients, and that strong mixing early on can wash out any gradients.

Recently, \citet{Ju2024} observationally examined this relation further, showing a connection between gas-phase metallicity gradients and ``diskiness" in $26$ galaxies at $z \sim 1$.
Their Galaxy ID9960 had the steepest metallicity gradient while also exhibiting a massive and regularly rotating disk, while Galaxy ID4391 had a complex (negative at small $R$ and positive at large $R$, averaging to roughly flat) metallicity gradient, while exhibiting a strongly turbulent disk.
This comparison offers key observational evidence of a relation between high (low) mixing in the ISM and weak (strong) metallicity radial gradients, in agreement with our work.

Finally, to our knowledge, none of the works that self-consistently model these various processes via cosmological simulations and that have argued that ISM metallicity radial gradients began strong and become shallower over time have shown that their trends for stars today agree with the MW, as we did for FIRE-2 in \citet{Graf2024}.
Furthermore, this represents the simplest interpretation of the MW results: that the age dependence of stellar metallicity gradients reflects the time evolution of the MW's ISM.
Constraints on ISM metallicity gradients from the MW and from galaxies across cosmic time provide critical benchmarks to inform not just metal-enrichment histories but also the evolution of turbulence and the nature of feedback in galaxies more generally.

\section{Acknowledgements}

RG and AW received support from the NSF via CAREER award AST-2045928 and grant AST-2107772.
We ran simulations using: XSEDE, supported by NSF grant ACI-1548562; Blue Waters, supported by the NSF; Frontera, supported by the NSF and TACC; Pleiades, via the NASA HEC program through the NAS Division at Ames Research Center.j

The data in these figures, and the Python code to generate them, are available at \url{https://github.com/rlgraf/Graf-et-al.-2024b.git} and archived in Zenodo \citep{zenodo}.
FIRE-2 simulations are publicly available \citep{Wetzel2023} at \url{http://flathub.flatironinstitute.org/fire}.
Additional FIRE simulation data is available at \url{http://fire.northwestern.edu/data}.
A public version of the \textsc{Gizmo} code is available at \url{www.tapir.caltech.edu/~phopkins/Site/GIZMO.html}.

\software{GizmoAnalysis \citep{GizmoAnalysis}}

\bibliographystyle{aasjournal}
\bibliography{article}{}

\begin{thebibliography}{}
\expandafter\ifx\csname natexlab\endcsname\relax\def\natexlab#1{#1}\fi
\providecommand{\url}[1]{\href{#1}{#1}}
\providecommand{\dodoi}[1]{doi:~\href{http://doi.org/#1}{\nolinkurl{#1}}}
\providecommand{\doeprint}[1]{\href{http://ascl.net/#1}{\nolinkurl{http://ascl.net/#1}}}
\providecommand{\doarXiv}[1]{\href{https://arxiv.org/abs/#1}{\nolinkurl{https://arxiv.org/abs/#1}}}

\bibitem[{{Acharyya} {et~al.}(2024){Acharyya}, {Peeples}, {Tumlinson}, {Shea}, {Lochhaas}, {Wright}, {Simons}, {Augustin}, {Smith}, \& {Hyeonmin Lee}}]{Acharyya2024}
{Acharyya}, A., {Peeples}, M.~S., {Tumlinson}, J., {et~al.} 2024, arXiv e-prints, arXiv:2404.06613, \dodoi{10.48550/arXiv.2404.06613}

\bibitem[{{Agertz} {et~al.}(2021){Agertz}, {Renaud}, {Feltzing}, {Read}, {Ryde}, {Andersson}, {Rey}, {Bensby}, \& {Feuillet}}]{Agertz2021}
{Agertz}, O., {Renaud}, F., {Feltzing}, S., {et~al.} 2021, \mnras, 503, 5826, \dodoi{10.1093/mnras/stab322}

\bibitem[{{Anders} {et~al.}(2023){Anders}, {Gispert}, {Ratcliffe}, {Chiappini}, {Minchev}, {Nepal}, {Queiroz}, {Amarante}, {Antoja}, {Casali}, {Casamiquela}, {Khalatyan}, {Miglio}, {Perottoni}, \& {Schultheis}}]{Anders2023}
{Anders}, F., {Gispert}, P., {Ratcliffe}, B., {et~al.} 2023, \aap, 678, A158, \dodoi{10.1051/0004-6361/202346666}

\bibitem[{{Asplund} {et~al.}(2009){Asplund}, {Grevesse}, {Sauval}, \& {Scott}}]{Asplund2009}
{Asplund}, M., {Grevesse}, N., {Sauval}, A.~J., \& {Scott}, P. 2009, \araa, 47, 481, \dodoi{10.1146/annurev.astro.46.060407.145222}

\bibitem[{{Belfiore} {et~al.}(2017){Belfiore}, {Maiolino}, {Tremonti}, {S{\'a}nchez}, {Bundy}, {Bershady}, {Westfall}, {Lin}, {Drory}, {Boquien}, {Thomas}, \& {Brinkmann}}]{Belfiore2017}
{Belfiore}, F., {Maiolino}, R., {Tremonti}, C., {et~al.} 2017, \mnras, 469, 151, \dodoi{10.1093/mnras/stx789}

\bibitem[{{Bellardini} {et~al.}(2022){Bellardini}, {Wetzel}, {Loebman}, \& {Bailin}}]{Bellardini2022}
{Bellardini}, M.~A., {Wetzel}, A., {Loebman}, S.~R., \& {Bailin}, J. 2022, \mnras, 514, 4270, \dodoi{10.1093/mnras/stac1637}

\bibitem[{{Bellardini} {et~al.}(2021){Bellardini}, {Wetzel}, {Loebman}, {Faucher-Gigu{\`e}re}, {Ma}, \& {Feldmann}}]{Bellardini2021}
{Bellardini}, M.~A., {Wetzel}, A., {Loebman}, S.~R., {et~al.} 2021, \mnras, 505, 4586, \dodoi{10.1093/mnras/stab1606}

\bibitem[{{Belokurov} \& {Kravtsov}(2022)}]{BelokurovKravtsov2022}
{Belokurov}, V., \& {Kravtsov}, A. 2022, \mnras, 514, 689, \dodoi{10.1093/mnras/stac1267}

\bibitem[{{Bird} {et~al.}(2012){Bird}, {Kazantzidis}, \& {Weinberg}}]{Bird2012}
{Bird}, J.~C., {Kazantzidis}, S., \& {Weinberg}, D.~H. 2012, \mnras, 420, 913, \dodoi{10.1111/j.1365-2966.2011.19728.x}

\bibitem[{{Bird} {et~al.}(2013){Bird}, {Kazantzidis}, {Weinberg}, {Guedes}, {Callegari}, {Mayer}, \& {Madau}}]{Bird2013}
{Bird}, J.~C., {Kazantzidis}, S., {Weinberg}, D.~H., {et~al.} 2013, \apj, 773, 43, \dodoi{10.1088/0004-637X/773/1/43}

\bibitem[{{Bird} {et~al.}(2021){Bird}, {Loebman}, {Weinberg}, {Brooks}, {Quinn}, \& {Christensen}}]{Bird2021}
{Bird}, J.~C., {Loebman}, S.~R., {Weinberg}, D.~H., {et~al.} 2021, \mnras, 503, 1815, \dodoi{10.1093/mnras/stab289}

\bibitem[{{Bland-Hawthorn} \& {Gerhard}(2016)}]{BlandHawthornGerhard2016}
{Bland-Hawthorn}, J., \& {Gerhard}, O. 2016, \araa, 54, 529, \dodoi{10.1146/annurev-astro-081915-023441}

\bibitem[{{Brook} {et~al.}(2012){Brook}, {Stinson}, {Gibson}, {Kawata}, {House}, {Miranda}, {Macci{\`o}}, {Pilkington}, {Ro{\v{s}}kar}, {Wadsley}, \& {Quinn}}]{Brook2012}
{Brook}, C.~B., {Stinson}, G.~S., {Gibson}, B.~K., {et~al.} 2012, \mnras, 426, 690, \dodoi{10.1111/j.1365-2966.2012.21738.x}

\bibitem[{{Buck} {et~al.}(2020){Buck}, {Obreja}, {Macci{\`o}}, {Minchev}, {Dutton}, \& {Ostriker}}]{Buck2020}
{Buck}, T., {Obreja}, A., {Macci{\`o}}, A.~V., {et~al.} 2020, \mnras, 491, 3461, \dodoi{10.1093/mnras/stz3241}

\bibitem[{{Buck} {et~al.}(2023){Buck}, {Obreja}, {Ratcliffe}, {Lu}, {Minchev}, \& {Macci{\`o}}}]{Buck2023}
{Buck}, T., {Obreja}, A., {Ratcliffe}, B., {et~al.} 2023, \mnras, 523, 1565, \dodoi{10.1093/mnras/stad1503}

\bibitem[{{Buitrago} \& {Trujillo}(2024)}]{Buitrago2024}
{Buitrago}, F., \& {Trujillo}, I. 2024, \aap, 682, A110, \dodoi{10.1051/0004-6361/202346133}

\bibitem[{{Carrillo} {et~al.}(2023){Carrillo}, {Ness}, {Hawkins}, {Sanderson}, {Wang}, {Wetzel}, \& {Bellardini}}]{Carillo2023}
{Carrillo}, A., {Ness}, M.~K., {Hawkins}, K., {et~al.} 2023, \apj, 942, 35, \dodoi{10.3847/1538-4357/aca1c7}

\bibitem[{{Cheng} {et~al.}(2024){Cheng}, {Giavalisco}, {Simons}, {Ji}, {Stroupe}, \& {Cleri}}]{Cheng2024}
{Cheng}, Y., {Giavalisco}, M., {Simons}, R.~C., {et~al.} 2024, arXiv e-prints, arXiv:2401.12319, \dodoi{10.48550/arXiv.2401.12319}

\bibitem[{{Chiappini} {et~al.}(2001){Chiappini}, {Matteucci}, \& {Romano}}]{Chiappini2001}
{Chiappini}, C., {Matteucci}, F., \& {Romano}, D. 2001, \apj, 554, 1044, \dodoi{10.1086/321427}

\bibitem[{{Conroy} {et~al.}(2022){Conroy}, {Weinberg}, {Naidu}, {Buck}, {Johnson}, {Cargile}, {Bonaca}, {Caldwell}, {Chandra}, {Han}, {Johnson}, {Speagle}, {Ting}, {Woody}, \& {Zaritsky}}]{Conroy2022}
{Conroy}, C., {Weinberg}, D.~H., {Naidu}, R.~P., {et~al.} 2022, arXiv e-prints, arXiv:2204.02989, \dodoi{10.48550/arXiv.2204.02989}

\bibitem[{{Curti} {et~al.}(2020){Curti}, {Maiolino}, {Cirasuolo}, {Mannucci}, {Williams}, {Auger}, {Mercurio}, {Hayden-Pawson}, {Cresci}, {Marconi}, {Belfiore}, {Cappellari}, {Cicone}, {Cullen}, {Meneghetti}, {Ota}, {Peng}, {Pettini}, {Swinbank}, \& {Troncoso}}]{Curti2020}
{Curti}, M., {Maiolino}, R., {Cirasuolo}, M., {et~al.} 2020, \mnras, 492, 821, \dodoi{10.1093/mnras/stz3379}

\bibitem[{{Donor} {et~al.}(2018){Donor}, {Frinchaboy}, {Cunha}, {Thompson}, {O'Connell}, {Zasowski}, {Jackson}, {Meyer McGrath}, {Almeida}, {Bizyaev}, {Carrera}, {Garc{\'\i}a-Hern{\'a}ndez}, {Nitschelm}, {Pan}, \& {Zamora}}]{Donor2018}
{Donor}, J., {Frinchaboy}, P.~M., {Cunha}, K., {et~al.} 2018, \aj, 156, 142, \dodoi{10.3847/1538-3881/aad635}

\bibitem[{{Escala} {et~al.}(2018){Escala}, {Wetzel}, {Kirby}, {Hopkins}, {Ma}, {Wheeler}, {Kere{\v{s}}}, {Faucher-Gigu{\`e}re}, \& {Quataert}}]{Escala2018}
{Escala}, I., {Wetzel}, A., {Kirby}, E.~N., {et~al.} 2018, \mnras, 474, 2194, \dodoi{10.1093/mnras/stx2858}

\bibitem[{{Faucher-Gigu{\`e}re} {et~al.}(2009){Faucher-Gigu{\`e}re}, {Lidz}, {Zaldarriaga}, \& {Hernquist}}]{FaucherGiguere2009}
{Faucher-Gigu{\`e}re}, C.-A., {Lidz}, A., {Zaldarriaga}, M., \& {Hernquist}, L. 2009, \apj, 703, 1416, \dodoi{10.1088/0004-637X/703/2/1416}

\bibitem[{{Frankel} {et~al.}(2019){Frankel}, {Sanders}, {Rix}, {Ting}, \& {Ness}}]{Frankel2019}
{Frankel}, N., {Sanders}, J., {Rix}, H.-W., {Ting}, Y.-S., \& {Ness}, M. 2019, \apj, 884, 99, \dodoi{10.3847/1538-4357/ab4254}

\bibitem[{{Freeman} \& {Bland-Hawthorn}(2002)}]{FreemanBlandHawthron2002}
{Freeman}, K., \& {Bland-Hawthorn}, J. 2002, \araa, 40, 487, \dodoi{10.1146/annurev.astro.40.060401.093840}

\bibitem[{{Gandhi} {et~al.}(2022){Gandhi}, {Wetzel}, {Hopkins}, {Shappee}, {Wheeler}, \& {Faucher-Gigu{\`e}re}}]{Gandhi2022}
{Gandhi}, P.~J., {Wetzel}, A., {Hopkins}, P.~F., {et~al.} 2022, \mnras, 516, 1941, \dodoi{10.1093/mnras/stac2228}

\bibitem[{{Garrison-Kimmel} {et~al.}(2017){Garrison-Kimmel}, {Wetzel}, {Bullock}, {Hopkins}, {Boylan-Kolchin}, {Faucher-Gigu{\`e}re}, {Kere{\v{s}}}, {Quataert}, {Sanderson}, {Graus}, \& {Kelley}}]{GarrisonKimmel2017}
{Garrison-Kimmel}, S., {Wetzel}, A., {Bullock}, J.~S., {et~al.} 2017, \mnras, 471, 1709, \dodoi{10.1093/mnras/stx1710}

\bibitem[{{Garrison-Kimmel} {et~al.}(2019{\natexlab{a}}){Garrison-Kimmel}, {Hopkins}, {Wetzel}, {Bullock}, {Boylan-Kolchin}, {Kere{\v{s}}}, {Faucher-Gigu{\`e}re}, {El-Badry}, {Lamberts}, {Quataert}, \& {Sanderson}}]{GarrisonKimmel2019a}
{Garrison-Kimmel}, S., {Hopkins}, P.~F., {Wetzel}, A., {et~al.} 2019{\natexlab{a}}, \mnras, 487, 1380, \dodoi{10.1093/mnras/stz1317}

\bibitem[{{Garrison-Kimmel} {et~al.}(2019{\natexlab{b}}){Garrison-Kimmel}, {Wetzel}, {Hopkins}, {Sanderson}, {El-Badry}, {Graus}, {Chan}, {Feldmann}, {Boylan-Kolchin}, {Hayward}, {Bullock}, {Fitts}, {Samuel}, {Wheeler}, {Kere{\v{s}}}, \& {Faucher-Gigu{\`e}re}}]{GarrisonKimmel2019b}
{Garrison-Kimmel}, S., {Wetzel}, A., {Hopkins}, P.~F., {et~al.} 2019{\natexlab{b}}, \mnras, 489, 4574, \dodoi{10.1093/mnras/stz2507}

\bibitem[{{Gibson} {et~al.}(2013){Gibson}, {Pilkington}, {Brook}, {Stinson}, \& {Bailin}}]{Gibson2013}
{Gibson}, B.~K., {Pilkington}, K., {Brook}, C.~B., {Stinson}, G.~S., \& {Bailin}, J. 2013, \aap, 554, A47, \dodoi{10.1051/0004-6361/201321239}

\bibitem[{Graf(2024)}]{zenodo}
Graf, R.~L. 2024, {The Data For Inside-Out versus Upside-Down: The Origin and Evolution of Metallicity Radial Gradients in FIRE Simulations of Milky Way-mass Galaxies and the Essential Role of Gas Mixing},  Zenodo, \dodoi{10.5281/zenodo.13992737}

\bibitem[{{Graf} {et~al.}(2024){Graf}, {Wetzel}, {Bellardini}, \& {Bailin}}]{Graf2024}
{Graf}, R.~L., {Wetzel}, A., {Bellardini}, M.~A., \& {Bailin}, J. 2024, arXiv e-prints, arXiv:2402.15614, \dodoi{10.48550/arXiv.2402.15614}

\bibitem[{{Gurvich} {et~al.}(2023){Gurvich}, {Stern}, {Faucher-Gigu{\`e}re}, {Hopkins}, {Wetzel}, {Moreno}, {Hayward}, {Richings}, \& {Hafen}}]{Gurvich2023}
{Gurvich}, A.~B., {Stern}, J., {Faucher-Gigu{\`e}re}, C.-A., {et~al.} 2023, \mnras, 519, 2598, \dodoi{10.1093/mnras/stac3712}

\bibitem[{{Hahn} \& {Abel}(2011)}]{HahnAbel2011}
{Hahn}, O., \& {Abel}, T. 2011, \mnras, 415, 2101, \dodoi{10.1111/j.1365-2966.2011.18820.x}

\bibitem[{{Han} {et~al.}(2024){Han}, {Yi}, {Oh}, {Pak}, {Croom}, {Devriendt}, {Dubois}, {Kimm}, {Kraljic}, {Pichon}, \& {Volonteri}}]{Han2024}
{Han}, S., {Yi}, S.~K., {Oh}, S., {et~al.} 2024, \apj, 968, 96, \dodoi{10.3847/1538-4357/ad43dc}

\bibitem[{{Hasheminia} {et~al.}(2022){Hasheminia}, {Mosleh}, {Tacchella}, {Hosseini-ShahiSavandi}, {Park}, \& {Naidu}}]{Hasheminia2022}
{Hasheminia}, M., {Mosleh}, M., {Tacchella}, S., {et~al.} 2022, \apjl, 932, L23, \dodoi{10.3847/2041-8213/ac76c8}

\bibitem[{{Hasheminia} {et~al.}(2024){Hasheminia}, {Mosleh}, {Zahra Hosseini-ShahiSavandi}, \& {Tacchella}}]{Hasheminia2024}
{Hasheminia}, M., {Mosleh}, M., {Zahra Hosseini-ShahiSavandi}, S., \& {Tacchella}, S. 2024, arXiv e-prints, arXiv:2410.05867, \dodoi{10.48550/arXiv.2410.05867}

\bibitem[{{Hawkins}(2023)}]{Hawkins2023}
{Hawkins}, K. 2023, \mnras, 525, 3318, \dodoi{10.1093/mnras/stad1244}

\bibitem[{{Hayward} \& {Hopkins}(2017)}]{Hayward2017}
{Hayward}, C.~C., \& {Hopkins}, P.~F. 2017, \mnras, 465, 1682, \dodoi{10.1093/mnras/stw2888}

\bibitem[{{Haywood}(2008)}]{Haywood2008}
{Haywood}, M. 2008, \mnras, 388, 1175, \dodoi{10.1111/j.1365-2966.2008.13395.x}

\bibitem[{{Hemler} {et~al.}(2021){Hemler}, {Torrey}, {Qi}, {Hernquist}, {Vogelsberger}, {Ma}, {Kewley}, {Nelson}, {Pillepich}, {Pakmor}, \& {Marinacci}}]{Hemler2021}
{Hemler}, Z.~S., {Torrey}, P., {Qi}, J., {et~al.} 2021, \mnras, 506, 3024, \dodoi{10.1093/mnras/stab1803}

\bibitem[{{Hernandez} {et~al.}(2021){Hernandez}, {Aloisi}, {James}, {Kumari}, {Berg}, {Adamo}, {Blair}, {Faucher-Gigu{\`e}re}, {Fox}, {Gurvich}, {Hafen}, {Heckman}, {Lebouteiller}, {Long}, {Skillman}, {Tumlinson}, \& {Whitmore}}]{Hernandez2021}
{Hernandez}, S., {Aloisi}, A., {James}, B.~L., {et~al.} 2021, \apj, 908, 226, \dodoi{10.3847/1538-4357/abd6c4}

\bibitem[{{Ho} {et~al.}(2015){Ho}, {Kudritzki}, {Kewley}, {Zahid}, {Dopita}, {Bresolin}, \& {Rupke}}]{Ho2015}
{Ho}, I.~T., {Kudritzki}, R.-P., {Kewley}, L.~J., {et~al.} 2015, \mnras, 448, 2030, \dodoi{10.1093/mnras/stv067}

\bibitem[{{Hopkins}(2015)}]{Hopkins2015}
{Hopkins}, P.~F. 2015, \mnras, 450, 53, \dodoi{10.1093/mnras/stv195}

\bibitem[{{Hopkins} {et~al.}(2018){Hopkins}, {Wetzel}, {Kere{\v{s}}}, {Faucher-Gigu{\`e}re}, {Quataert}, {Boylan-Kolchin}, {Murray}, {Hayward}, {Garrison-Kimmel}, {Hummels}, {Feldmann}, {Torrey}, {Ma}, {Angl{\'e}s-Alc{\'a}zar}, {Su}, {Orr}, {Schmitz}, {Escala}, {Sanderson}, {Grudi{\'c}}, {Hafen}, {Kim}, {Fitts}, {Bullock}, {Wheeler}, {Chan}, {Elbert}, \& {Narayanan}}]{Hopkins2018}
{Hopkins}, P.~F., {Wetzel}, A., {Kere{\v{s}}}, D., {et~al.} 2018, \mnras, 480, 800, \dodoi{10.1093/mnras/sty1690}

\bibitem[{{Hopkins} {et~al.}(2020){Hopkins}, {Chan}, {Garrison-Kimmel}, {Ji}, {Su}, {Hummels}, {Kere{\v{s}}}, {Quataert}, \& {Faucher-Gigu{\`e}re}}]{Hopkins2020}
{Hopkins}, P.~F., {Chan}, T.~K., {Garrison-Kimmel}, S., {et~al.} 2020, \mnras, 492, 3465, \dodoi{10.1093/mnras/stz3321}

\bibitem[{{Hopkins} {et~al.}(2023){Hopkins}, {Wetzel}, {Wheeler}, {Sanderson}, {Grudi{\'c}}, {Sameie}, {Boylan-Kolchin}, {Orr}, {Ma}, {Faucher-Gigu{\`e}re}, {Kere{\v{s}}}, {Quataert}, {Su}, {Moreno}, {Feldmann}, {Bullock}, {Loebman}, {Angl{\'e}s-Alc{\'a}zar}, {Stern}, {Necib}, {Choban}, \& {Hayward}}]{Hopkins2023}
{Hopkins}, P.~F., {Wetzel}, A., {Wheeler}, C., {et~al.} 2023, \mnras, 519, 3154, \dodoi{10.1093/mnras/stac3489}

\bibitem[{{Ibarra-Medel} {et~al.}(2016){Ibarra-Medel}, {S{\'a}nchez}, {Avila-Reese}, {Hern{\'a}ndez-Toledo}, {Gonz{\'a}lez}, {Drory}, {Bundy}, {Bizyaev}, {Cano-D{\'\i}az}, {Malanushenko}, {Pan}, {Roman-Lopes}, \& {Thomas}}]{IbarraMedel2016}
{Ibarra-Medel}, H.~J., {S{\'a}nchez}, S.~F., {Avila-Reese}, V., {et~al.} 2016, \mnras, 463, 2799, \dodoi{10.1093/mnras/stw2126}

\bibitem[{{Imig} {et~al.}(2023){Imig}, {Price}, {Holtzman}, {Stone-Martinez}, {Majewski}, {Weinberg}, {Johnson}, {Allende Prieto}, {Beaton}, {Beers}, {Bizyaev}, {Blanton}, {Brownstein}, {Cunha}, {Fern{\'a}ndez-Trincado}, {Feuillet}, {Hasselquist}, {Hayes}, {J{\"o}nsson}, {Lane}, {Lian}, {M{\'e}sz{\'a}ros}, {Nidever}, {Robin}, {Shetrone}, {Smith}, \& {Wilson}}]{Imig2023}
{Imig}, J., {Price}, C., {Holtzman}, J.~A., {et~al.} 2023, \apj, 954, 124, \dodoi{10.3847/1538-4357/ace9b8}

\bibitem[{{Irodotou} {et~al.}(2022){Irodotou}, {Fragkoudi}, {Pakmor}, {Grand}, {Gadotti}, {Costa}, {Springel}, {G{\'o}mez}, \& {Marinacci}}]{Irodotou2022}
{Irodotou}, D., {Fragkoudi}, F., {Pakmor}, R., {et~al.} 2022, \mnras, 513, 3768, \dodoi{10.1093/mnras/stac1143}

\bibitem[{{Iwamoto} {et~al.}(1999){Iwamoto}, {Brachwitz}, {Nomoto}, {Kishimoto}, {Umeda}, {Hix}, \& {Thielemann}}]{Iwamoto1999}
{Iwamoto}, K., {Brachwitz}, F., {Nomoto}, K., {et~al.} 1999, \apjs, 125, 439, \dodoi{10.1086/313278}

\bibitem[{{Iza} {et~al.}(2024){Iza}, {Nuza}, {Scannapieco}, {Grand}, {G{\'o}mez}, {Springel}, {Pakmor}, {Marinacci}, \& {Fragkoudi}}]{Iza2024}
{Iza}, F.~G., {Nuza}, S.~E., {Scannapieco}, C., {et~al.} 2024, \mnras, 528, 1737, \dodoi{10.1093/mnras/stae110}

\bibitem[{{Izzard} {et~al.}(2004){Izzard}, {Tout}, {Karakas}, \& {Pols}}]{Izzard2004}
{Izzard}, R.~G., {Tout}, C.~A., {Karakas}, A.~I., \& {Pols}, O.~R. 2004, \mnras, 350, 407, \dodoi{10.1111/j.1365-2966.2004.07446.x}

\bibitem[{{Jones} {et~al.}(2013){Jones}, {Ellis}, {Richard}, \& {Jullo}}]{Jones2013}
{Jones}, T., {Ellis}, R.~S., {Richard}, J., \& {Jullo}, E. 2013, \apj, 765, 48, \dodoi{10.1088/0004-637X/765/1/48}

\bibitem[{{Jones} {et~al.}(2018){Jones}, {Stark}, \& {Ellis}}]{Jones2018}
{Jones}, T., {Stark}, D.~P., \& {Ellis}, R.~S. 2018, \apj, 863, 191, \dodoi{10.3847/1538-4357/aad37f}

\bibitem[{{Ju} {et~al.}(2024){Ju}, {Wang}, {Jones}, {Bari{\v{s}}i{\'c}}, {Nanayakkara}, {Bundy}, {Faucher-Gigu{\`e}re}, {Feng}, {Glazebrook}, {Henry}, {Malkan}, {Obreschkow}, {Roy}, {Sanders}, {Sun}, \& {Treu}}]{Ju2024}
{Ju}, M., {Wang}, X., {Jones}, T., {et~al.} 2024, arXiv e-prints, arXiv:2409.01616.
\newblock \doarXiv{2409.01616}

\bibitem[{{Kassin} {et~al.}(2012){Kassin}, {Weiner}, {Faber}, {Gardner}, {Willmer}, {Coil}, {Cooper}, {Devriendt}, {Dutton}, {Guhathakurta}, {Koo}, {Metevier}, {Noeske}, \& {Primack}}]{Kassin2012}
{Kassin}, S.~A., {Weiner}, B.~J., {Faber}, S.~M., {et~al.} 2012, \apj, 758, 106, \dodoi{10.1088/0004-637X/758/2/106}

\bibitem[{{Kennicutt}(1998)}]{Kennicutt1998}
{Kennicutt}, Robert~C., J. 1998, \araa, 36, 189, \dodoi{10.1146/annurev.astro.36.1.189}

\bibitem[{{Kreckel} {et~al.}(2019){Kreckel}, {Ho}, {Blanc}, {Groves}, {Santoro}, {Schinnerer}, {Bigiel}, {Chevance}, {Congiu}, {Emsellem}, {Faesi}, {Glover}, {Grasha}, {Kruijssen}, {Lang}, {Leroy}, {Meidt}, {McElroy}, {Pety}, {Rosolowsky}, {Saito}, {Sandstrom}, {Sanchez-Blazquez}, \& {Schruba}}]{Kreckel2019}
{Kreckel}, K., {Ho}, I.~T., {Blanc}, G.~A., {et~al.} 2019, \apj, 887, 80, \dodoi{10.3847/1538-4357/ab5115}

\bibitem[{{Kreckel} {et~al.}(2020){Kreckel}, {Ho}, {Blanc}, {Glover}, {Groves}, {Rosolowsky}, {Bigiel}, {Boqu{\'\i}en}, {Chevance}, {Dale}, {Deger}, {Emsellem}, {Grasha}, {Kim}, {Klessen}, {Kruijssen}, {Lee}, {Leroy}, {Liu}, {McElroy}, {Meidt}, {Pessa}, {Sanchez-Blazquez}, {Sandstrom}, {Santoro}, {Scheuermann}, {Schinnerer}, {Schruba}, {Utomo}, {Watkins}, \& {Williams}}]{Kreckel2020}
---. 2020, \mnras, 499, 193, \dodoi{10.1093/mnras/staa2743}

\bibitem[{{Kroupa}(2001)}]{Kroupa2001}
{Kroupa}, P. 2001, \mnras, 322, 231, \dodoi{10.1046/j.1365-8711.2001.04022.x}

\bibitem[{{Krumholz} \& {Gnedin}(2011)}]{KrumholzGnedin2011}
{Krumholz}, M.~R., \& {Gnedin}, N.~Y. 2011, \apj, 729, 36, \dodoi{10.1088/0004-637X/729/1/36}

\bibitem[{{Leitherer} {et~al.}(1999){Leitherer}, {Schaerer}, {Goldader}, {Delgado}, {Robert}, {Kune}, {de Mello}, {Devost}, \& {Heckman}}]{Leitherer1999}
{Leitherer}, C., {Schaerer}, D., {Goldader}, J.~D., {et~al.} 1999, \apjs, 123, 3, \dodoi{10.1086/313233}

\bibitem[{{Lian} {et~al.}(2023){Lian}, {Bergemann}, {Pillepich}, {Zasowski}, \& {Lane}}]{Lian2023}
{Lian}, J., {Bergemann}, M., {Pillepich}, A., {Zasowski}, G., \& {Lane}, R.~R. 2023, Nature Astronomy, 7, 951, \dodoi{10.1038/s41550-023-01977-z}

\bibitem[{{Lian} {et~al.}(2024){Lian}, {Zasowski}, {Chen}, {Imig}, {Wang}, {Boardman}, \& {Liu}}]{Lian2024}
{Lian}, J., {Zasowski}, G., {Chen}, B., {et~al.} 2024, Nature Astronomy, \dodoi{10.1038/s41550-024-02315-7}

\bibitem[{{Lian} {et~al.}(2022){Lian}, {Zasowski}, {Hasselquist}, {Holtzman}, {Boardman}, {Cunha}, {Fern{\'a}ndez-Trincado}, {Frinchaboy}, {Garcia-Hernandez}, {Nitschelm}, {Lane}, {Thomas}, \& {Zhang}}]{Lian2022}
{Lian}, J., {Zasowski}, G., {Hasselquist}, S., {et~al.} 2022, \mnras, 511, 5639, \dodoi{10.1093/mnras/stac479}

\bibitem[{{Lu} {et~al.}(2022){Lu}, {Buck}, {Minchev}, \& {Ness}}]{Lu2022b}
{Lu}, Y., {Buck}, T., {Minchev}, I., \& {Ness}, M.~K. 2022, \mnras, 515, L34, \dodoi{10.1093/mnrasl/slac065}

\bibitem[{{Ma} {et~al.}(2017{\natexlab{a}}){Ma}, {Hopkins}, {Feldmann}, {Torrey}, {Faucher-Gigu{\`e}re}, \& {Kere{\v{s}}}}]{Ma2017a}
{Ma}, X., {Hopkins}, P.~F., {Feldmann}, R., {et~al.} 2017{\natexlab{a}}, \mnras, 466, 4780, \dodoi{10.1093/mnras/stx034}

\bibitem[{{Ma} {et~al.}(2017{\natexlab{b}}){Ma}, {Hopkins}, {Wetzel}, {Kirby}, {Angl{\'e}s-Alc{\'a}zar}, {Faucher-Gigu{\`e}re}, {Kere{\v{s}}}, \& {Quataert}}]{Ma2017b}
{Ma}, X., {Hopkins}, P.~F., {Wetzel}, A.~R., {et~al.} 2017{\natexlab{b}}, \mnras, 467, 2430, \dodoi{10.1093/mnras/stx273}

\bibitem[{{Mannucci} {et~al.}(2006){Mannucci}, {Della Valle}, \& {Panagia}}]{Mannucci2006}
{Mannucci}, F., {Della Valle}, M., \& {Panagia}, N. 2006, \mnras, 370, 773, \dodoi{10.1111/j.1365-2966.2006.10501.x}

\bibitem[{{Marigo}(2001)}]{Marigo2001}
{Marigo}, P. 2001, \aap, 370, 194, \dodoi{10.1051/0004-6361:20000247}

\bibitem[{{McCluskey} {et~al.}(2024){McCluskey}, {Wetzel}, {Loebman}, {Moreno}, {Faucher-Gigu{\`e}re}, \& {Hopkins}}]{McCluskey2023}
{McCluskey}, F., {Wetzel}, A., {Loebman}, S.~R., {et~al.} 2024, \mnras, 527, 6926, \dodoi{10.1093/mnras/stad3547}

\bibitem[{{Mercedes-Feliz} {et~al.}(2023){Mercedes-Feliz}, {Angl{\'e}s-Alc{\'a}zar}, {Hayward}, {Cochrane}, {Terrazas}, {Wellons}, {Richings}, {Faucher-Gigu{\`e}re}, {Moreno}, {Su}, {Hopkins}, {Quataert}, \& {Kere{\v{s}}}}]{MercedesFeliz2023}
{Mercedes-Feliz}, J., {Angl{\'e}s-Alc{\'a}zar}, D., {Hayward}, C.~C., {et~al.} 2023, \mnras, 524, 3446, \dodoi{10.1093/mnras/stad2079}

\bibitem[{{Minchev} {et~al.}(2013){Minchev}, {Chiappini}, \& {Martig}}]{Minchev2013}
{Minchev}, I., {Chiappini}, C., \& {Martig}, M. 2013, \aap, 558, A9, \dodoi{10.1051/0004-6361/201220189}

\bibitem[{{Minchev} {et~al.}(2018){Minchev}, {Anders}, {Recio-Blanco}, {Chiappini}, {de Laverny}, {Queiroz}, {Steinmetz}, {Adibekyan}, {Carrillo}, {Cescutti}, {Guiglion}, {Hayden}, {de Jong}, {Kordopatis}, {Majewski}, {Martig}, \& {Santiago}}]{Minchev2018}
{Minchev}, I., {Anders}, F., {Recio-Blanco}, A., {et~al.} 2018, \mnras, 481, 1645, \dodoi{10.1093/mnras/sty2033}

\bibitem[{{Moll{\'a}} {et~al.}(2019){Moll{\'a}}, {D{\'\i}az}, {Cavichia}, {Gibson}, {Maciel}, {Costa}, {Ascasibar}, \& {Few}}]{Molla2019}
{Moll{\'a}}, M., {D{\'\i}az}, {\'A}.~I., {Cavichia}, O., {et~al.} 2019, \mnras, 482, 3071, \dodoi{10.1093/mnras/sty2877}

\bibitem[{{Muley} {et~al.}(2021){Muley}, {Wheeler}, {Hopkins}, {Wetzel}, {Emerick}, \& {Kere{\v{s}}}}]{Muley2021}
{Muley}, D.~A., {Wheeler}, C.~R., {Hopkins}, P.~F., {et~al.} 2021, \mnras, 508, 508, \dodoi{10.1093/mnras/stab2572}

\bibitem[{{Mun} {et~al.}(2024){Mun}, {Wisnioski}, {Battisti}, {Mendel}, {Ellison}, {Taylor}, {Lagos}, {Harborne}, {Foster}, {Croom}, {Bellstedt}, {Barsanti}, {Gupta}, {Valenzuela}, {Chen}, {Grasha}, {Mukherjee}, {Park}, {Sharda}, {Sweet}, {Remus}, \& {Zafar}}]{Mun2024}
{Mun}, M., {Wisnioski}, E., {Battisti}, A.~J., {et~al.} 2024, \mnras, 530, 5072, \dodoi{10.1093/mnras/stae1132}

\bibitem[{{Nomoto} {et~al.}(2006){Nomoto}, {Tominaga}, {Umeda}, {Kobayashi}, \& {Maeda}}]{Nomoto2006}
{Nomoto}, K., {Tominaga}, N., {Umeda}, H., {Kobayashi}, C., \& {Maeda}, K. 2006, \nphysa, 777, 424, \dodoi{10.1016/j.nuclphysa.2006.05.008}

\bibitem[{{Okalidis} {et~al.}(2022){Okalidis}, {Grand}, {Yates}, \& {Springel}}]{Okalidis2022}
{Okalidis}, P., {Grand}, R. J.~J., {Yates}, R.~M., \& {Springel}, V. 2022, \mnras, 514, 5085, \dodoi{10.1093/mnras/stac1635}

\bibitem[{{Orr} {et~al.}(2022{\natexlab{a}}){Orr}, {Fielding}, {Hayward}, \& {Burkhart}}]{Orr2022a}
{Orr}, M.~E., {Fielding}, D.~B., {Hayward}, C.~C., \& {Burkhart}, B. 2022{\natexlab{a}}, \apjl, 924, L28, \dodoi{10.3847/2041-8213/ac479f}

\bibitem[{{Orr} {et~al.}(2022{\natexlab{b}}){Orr}, {Fielding}, {Hayward}, \& {Burkhart}}]{Orr2022b}
---. 2022{\natexlab{b}}, \apj, 932, 88, \dodoi{10.3847/1538-4357/ac6c26}

\bibitem[{{Orr} {et~al.}(2018){Orr}, {Hayward}, {Hopkins}, {Chan}, {Faucher-Gigu{\`e}re}, {Feldmann}, {Kere{\v{s}}}, {Murray}, \& {Quataert}}]{Orr2018}
{Orr}, M.~E., {Hayward}, C.~C., {Hopkins}, P.~F., {et~al.} 2018, \mnras, 478, 3653, \dodoi{10.1093/mnras/sty1241}

\bibitem[{{Orr} {et~al.}(2023){Orr}, {Burkhart}, {Wetzel}, {Hopkins}, {Escala}, {Strom}, {Goldsmith}, {Pineda}, {Hayward}, \& {Loebman}}]{Orr2023}
{Orr}, M.~E., {Burkhart}, B., {Wetzel}, A., {et~al.} 2023, \mnras, 521, 3708, \dodoi{10.1093/mnras/stad676}

\bibitem[{{Pan} {et~al.}(2023){Pan}, {Zheng}, \& {Kong}}]{Pan2023}
{Pan}, Z., {Zheng}, X., \& {Kong}, X. 2023, \apj, 958, 42, \dodoi{10.3847/1538-4357/ad0230}

\bibitem[{{Pandya} {et~al.}(2021){Pandya}, {Fielding}, {Angl{\'e}s-Alc{\'a}zar}, {Somerville}, {Bryan}, {Hayward}, {Stern}, {Kim}, {Quataert}, {Forbes}, {Faucher-Gigu{\`e}re}, {Feldmann}, {Hafen}, {Hopkins}, {Kere{\v{s}}}, {Murray}, \& {Wetzel}}]{Pandya2021}
{Pandya}, V., {Fielding}, D.~B., {Angl{\'e}s-Alc{\'a}zar}, D., {et~al.} 2021, \mnras, 508, 2979, \dodoi{10.1093/mnras/stab2714}

\bibitem[{{Pilkington} {et~al.}(2012){Pilkington}, {Few}, {Gibson}, {Calura}, {Michel-Dansac}, {Thacker}, {Moll{\'a}}, {Matteucci}, {Rahimi}, {Kawata}, {Kobayashi}, {Brook}, {Stinson}, {Couchman}, {Bailin}, \& {Wadsley}}]{Pilkington2012}
{Pilkington}, K., {Few}, C.~G., {Gibson}, B.~K., {et~al.} 2012, \aap, 540, A56, \dodoi{10.1051/0004-6361/201117466}

\bibitem[{{Planck Collaboration} {et~al.}(2020){Planck Collaboration}, {Aghanim}, {Akrami}, {Ashdown}, {Aumont}, {Baccigalupi}, {Ballardini}, {Banday}, {Barreiro}, {Bartolo}, {Basak}, {Battye}, {Benabed}, {Bernard}, {Bersanelli}, {Bielewicz}, {Bock}, {Bond}, {Borrill}, {Bouchet}, {Boulanger}, {Bucher}, {Burigana}, {Butler}, {Calabrese}, {Cardoso}, {Carron}, {Challinor}, {Chiang}, {Chluba}, {Colombo}, {Combet}, {Contreras}, {Crill}, {Cuttaia}, {de Bernardis}, {de Zotti}, {Delabrouille}, {Delouis}, {Di Valentino}, {Diego}, {Dor{\'e}}, {Douspis}, {Ducout}, {Dupac}, {Dusini}, {Efstathiou}, {Elsner}, {En{\ss}lin}, {Eriksen}, {Fantaye}, {Farhang}, {Fergusson}, {Fernandez-Cobos}, {Finelli}, {Forastieri}, {Frailis}, {Fraisse}, {Franceschi}, {Frolov}, {Galeotta}, {Galli}, {Ganga}, {G{\'e}nova-Santos}, {Gerbino}, {Ghosh}, {Gonz{\'a}lez-Nuevo}, {G{\'o}rski}, {Gratton}, {Gruppuso}, {Gudmundsson}, {Hamann}, {Handley}, {Hansen}, {Herranz}, {Hildebrandt}, {Hivon}, {Huang}, {Jaffe}, {Jones}, {Karakci}, {Keih{\"a}nen},
  {Keskitalo}, {Kiiveri}, {Kim}, {Kisner}, {Knox}, {Krachmalnicoff}, {Kunz}, {Kurki-Suonio}, {Lagache}, {Lamarre}, {Lasenby}, {Lattanzi}, {Lawrence}, {Le Jeune}, {Lemos}, {Lesgourgues}, {Levrier}, {Lewis}, {Liguori}, {Lilje}, {Lilley}, {Lindholm}, {L{\'o}pez-Caniego}, {Lubin}, {Ma}, {Mac{\'\i}as-P{\'e}rez}, {Maggio}, {Maino}, {Mandolesi}, {Mangilli}, {Marcos-Caballero}, {Maris}, {Martin}, {Martinelli}, {Mart{\'\i}nez-Gonz{\'a}lez}, {Matarrese}, {Mauri}, {McEwen}, {Meinhold}, {Melchiorri}, {Mennella}, {Migliaccio}, {Millea}, {Mitra}, {Miville-Desch{\^e}nes}, {Molinari}, {Montier}, {Morgante}, {Moss}, {Natoli}, {N{\o}rgaard-Nielsen}, {Pagano}, {Paoletti}, {Partridge}, {Patanchon}, {Peiris}, {Perrotta}, {Pettorino}, {Piacentini}, {Polastri}, {Polenta}, {Puget}, {Rachen}, {Reinecke}, {Remazeilles}, {Renzi}, {Rocha}, {Rosset}, {Roudier}, {Rubi{\~n}o-Mart{\'\i}n}, {Ruiz-Granados}, {Salvati}, {Sandri}, {Savelainen}, {Scott}, {Shellard}, {Sirignano}, {Sirri}, {Spencer}, {Sunyaev}, {Suur-Uski}, {Tauber}, {Tavagnacco},
  {Tenti}, {Toffolatti}, {Tomasi}, {Trombetti}, {Valenziano}, {Valiviita}, {Van Tent}, {Vibert}, {Vielva}, {Villa}, {Vittorio}, {Wandelt}, {Wehus}, {White}, {White}, {Zacchei}, \& {Zonca}}]{PlanckCollab2020}
{Planck Collaboration}, {Aghanim}, N., {Akrami}, Y., {et~al.} 2020, \aap, 641, A6, \dodoi{10.1051/0004-6361/201833910}

\bibitem[{{Ratcliffe} {et~al.}(2023){Ratcliffe}, {Minchev}, {Anders}, {Khoperskov}, {Guiglion}, {Buck}, {Cunha}, {Queiroz}, {Nitschelm}, {Meszaros}, {Steinmetz}, {de Jong}, {Nepal}, {Lane}, \& {Sobeck}}]{Ratcliffe2023}
{Ratcliffe}, B., {Minchev}, I., {Anders}, F., {et~al.} 2023, \mnras, 525, 2208, \dodoi{10.1093/mnras/stad1573}

\bibitem[{{Renaud} {et~al.}(2024){Renaud}, {Ratcliffe}, {Minchev}, {Haywood}, {Di Matteo}, {Agertz}, \& {Romeo}}]{Renaud2024}
{Renaud}, F., {Ratcliffe}, B., {Minchev}, I., {et~al.} 2024, arXiv e-prints, arXiv:2409.10598, \dodoi{10.48550/arXiv.2409.10598}

\bibitem[{{Rizzo} {et~al.}(2020){Rizzo}, {Vegetti}, {Powell}, {Fraternali}, {McKean}, {Stacey}, \& {White}}]{Rizzo2020}
{Rizzo}, F., {Vegetti}, S., {Powell}, D., {et~al.} 2020, \nat, 584, 201, \dodoi{10.1038/s41586-020-2572-6}

\bibitem[{{Saintonge} \& {Catinella}(2022)}]{SaintongeCatinella2022}
{Saintonge}, A., \& {Catinella}, B. 2022, \araa, 60, 319, \dodoi{10.1146/annurev-astro-021022-043545}

\bibitem[{{Sakhibov} {et~al.}(2018){Sakhibov}, {Zinchenko}, {Pilyugin}, {Grebel}, {Just}, \& {V{\'\i}lchez}}]{Sakhibov2018}
{Sakhibov}, F., {Zinchenko}, I.~A., {Pilyugin}, L.~S., {et~al.} 2018, \mnras, 474, 1657, \dodoi{10.1093/mnras/stx2799}

\bibitem[{{S{\'a}nchez-Menguiano} {et~al.}(2016){S{\'a}nchez-Menguiano}, {S{\'a}nchez}, {P{\'e}rez}, {Garc{\'\i}a-Benito}, {Husemann}, {Mast}, {Mendoza}, {Ruiz-Lara}, {Ascasibar}, {Bland-Hawthorn}, {Cavichia}, {D{\'\i}az}, {Florido}, {Galbany}, {G{\'o}nzalez Delgado}, {Kehrig}, {Marino}, {M{\'a}rquez}, {Masegosa}, {M{\'e}ndez-Abreu}, {Moll{\'a}}, {Del Olmo}, {P{\'e}rez}, {S{\'a}nchez-Bl{\'a}zquez}, {Stanishev}, {Walcher}, {L{\'o}pez-S{\'a}nchez}, \& {CALIFA Collaboration}}]{SanchezMenguiano2016}
{S{\'a}nchez-Menguiano}, L., {S{\'a}nchez}, S.~F., {P{\'e}rez}, I., {et~al.} 2016, \aap, 587, A70, \dodoi{10.1051/0004-6361/201527450}

\bibitem[{{Sanders} {et~al.}(2012){Sanders}, {Caldwell}, {McDowell}, \& {Harding}}]{Sanders2012}
{Sanders}, N.~E., {Caldwell}, N., {McDowell}, J., \& {Harding}, P. 2012, \apj, 758, 133, \dodoi{10.1088/0004-637X/758/2/133}

\bibitem[{{Santistevan} {et~al.}(2020){Santistevan}, {Wetzel}, {El-Badry}, {Bland-Hawthorn}, {Boylan-Kolchin}, {Bailin}, {Faucher-Gigu{\`e}re}, \& {Benincasa}}]{Santistevan2020}
{Santistevan}, I.~B., {Wetzel}, A., {El-Badry}, K., {et~al.} 2020, \mnras, 497, 747, \dodoi{10.1093/mnras/staa1923}

\bibitem[{{Schmidt}(1959)}]{Schmidt1959}
{Schmidt}, M. 1959, \apj, 129, 243, \dodoi{10.1086/146614}

\bibitem[{{Searle}(1971)}]{Searle1971}
{Searle}, L. 1971, \apj, 168, 327, \dodoi{10.1086/151090}

\bibitem[{{Sellwood} \& {Binney}(2002)}]{SellwoodBinney2002}
{Sellwood}, J.~A., \& {Binney}, J.~J. 2002, \mnras, 336, 785, \dodoi{10.1046/j.1365-8711.2002.05806.x}

\bibitem[{{Sharda} {et~al.}(2021){Sharda}, {Krumholz}, {Wisnioski}, {Forbes}, {Federrath}, \& {Acharyya}}]{Sharda2021}
{Sharda}, P., {Krumholz}, M.~R., {Wisnioski}, E., {et~al.} 2021, \mnras, 502, 5935, \dodoi{10.1093/mnras/stab252}

\bibitem[{{Springel}(2005)}]{Springel2005}
{Springel}, V. 2005, \mnras, 364, 1105, \dodoi{10.1111/j.1365-2966.2005.09655.x}

\bibitem[{{Su} {et~al.}(2017){Su}, {Hopkins}, {Hayward}, {Faucher-Gigu{\`e}re}, {Kere{\v{s}}}, {Ma}, \& {Robles}}]{Su2017}
{Su}, K.-Y., {Hopkins}, P.~F., {Hayward}, C.~C., {et~al.} 2017, \mnras, 471, 144, \dodoi{10.1093/mnras/stx1463}

\bibitem[{{Sun} {et~al.}(2024){Sun}, {Wang}, {Ma}, {Wang}, {Wetzel}, {Faucher-Gigu{\`e}re}, {Hopkins}, {Kere{\v{s}}}, {Graf}, {Marszewski}, {Stern}, {Sun}, {Sun}, \& {Thyme}}]{Sun2024}
{Sun}, X., {Wang}, X., {Ma}, X., {et~al.} 2024, arXiv e-prints, arXiv:2409.09290, \dodoi{10.48550/arXiv.2409.09290}

\bibitem[{{Tan} {et~al.}(2024){Tan}, {Muzzin}, {Marchesini}, {Sok}, {Sarrouh}, \& {Marsan}}]{Tan2024}
{Tan}, V. Y.~Y., {Muzzin}, A., {Marchesini}, D., {et~al.} 2024, \apj, 964, 177, \dodoi{10.3847/1538-4357/ad2c90}

\bibitem[{{Tissera} {et~al.}(2022){Tissera}, {Rosas-Guevara}, {Sillero}, {Pedrosa}, {Theuns}, \& {Bignone}}]{Tissera2022}
{Tissera}, P.~B., {Rosas-Guevara}, Y., {Sillero}, E., {et~al.} 2022, \mnras, 511, 1667, \dodoi{10.1093/mnras/stab3644}

\bibitem[{{Trapp} {et~al.}(2022){Trapp}, {Kere{\v{s}}}, {Chan}, {Escala}, {Hummels}, {Hopkins}, {Faucher-Gigu{\`e}re}, {Murray}, {Quataert}, \& {Wetzel}}]{Trapp2022}
{Trapp}, C.~W., {Kere{\v{s}}}, D., {Chan}, T.~K., {et~al.} 2022, \mnras, 509, 4149, \dodoi{10.1093/mnras/stab3251}

\bibitem[{{{\"U}bler} {et~al.}(2019){{\"U}bler}, {Genzel}, {Wisnioski}, {F{\"o}rster Schreiber}, {Shimizu}, {Price}, {Tacconi}, {Belli}, {Wilman}, {Fossati}, {Mendel}, {Davies}, {Beifiori}, {Bender}, {Brammer}, {Burkert}, {Chan}, {Davies}, {Fabricius}, {Galametz}, {Herrera-Camus}, {Lang}, {Lutz}, {Momcheva}, {Naab}, {Nelson}, {Saglia}, {Tadaki}, {van Dokkum}, \& {Wuyts}}]{Ubler2019}
{{\"U}bler}, H., {Genzel}, R., {Wisnioski}, E., {et~al.} 2019, \apj, 880, 48, \dodoi{10.3847/1538-4357/ab27cc}

\bibitem[{{van den Hoek} \& {Groenewegen}(1997)}]{VandenHoek1997}
{van den Hoek}, L.~B., \& {Groenewegen}, M.~A.~T. 1997, \aaps, 123, 305, \dodoi{10.1051/aas:1997162}

\bibitem[{{Vickers} {et~al.}(2021){Vickers}, {Shen}, \& {Li}}]{Vickers2021}
{Vickers}, J.~J., {Shen}, J., \& {Li}, Z.-Y. 2021, \apj, 922, 189, \dodoi{10.3847/1538-4357/ac27a9}

\bibitem[{{Vincenzo} \& {Kobayashi}(2018)}]{VK2018}
{Vincenzo}, F., \& {Kobayashi}, C. 2018, \mnras, 478, 155, \dodoi{10.1093/mnras/sty1047}

\bibitem[{{Vincenzo} \& {Kobayashi}(2020)}]{VK2020}
---. 2020, \mnras, 496, 80, \dodoi{10.1093/mnras/staa1451}

\bibitem[{{Wang} {et~al.}(2024){Wang}, {Carrillo}, {Ness}, \& {Buck}}]{Wang2024}
{Wang}, K., {Carrillo}, A., {Ness}, M.~K., \& {Buck}, T. 2024, \mnras, 527, 321, \dodoi{10.1093/mnras/stad3182}

\bibitem[{{Wang} {et~al.}(2022){Wang}, {Jones}, {Vulcani}, {Treu}, {Morishita}, {Roberts-Borsani}, {Malkan}, {Henry}, {Brammer}, {Strait}, {Brada{\v{c}}}, {Boyett}, {Calabr{\`o}}, {Castellano}, {Fontana}, {Glazebrook}, {Kelly}, {Leethochawalit}, {Marchesini}, {Santini}, {Trenti}, \& {Yang}}]{Wang2022}
{Wang}, X., {Jones}, T., {Vulcani}, B., {et~al.} 2022, \apjl, 938, L16, \dodoi{10.3847/2041-8213/ac959e}

\bibitem[{{Wellons} {et~al.}(2023){Wellons}, {Faucher-Gigu{\`e}re}, {Hopkins}, {Quataert}, {Angl{\'e}s-Alc{\'a}zar}, {Feldmann}, {Hayward}, {Kere{\v{s}}}, {Su}, \& {Wetzel}}]{Wellons2023}
{Wellons}, S., {Faucher-Gigu{\`e}re}, C.-A., {Hopkins}, P.~F., {et~al.} 2023, \mnras, 520, 5394, \dodoi{10.1093/mnras/stad511}

\bibitem[{{Wenger} {et~al.}(2019){Wenger}, {Balser}, {Anderson}, \& {Bania}}]{Wenger2019}
{Wenger}, T.~V., {Balser}, D.~S., {Anderson}, L.~D., \& {Bania}, T.~M. 2019, \apj, 887, 114, \dodoi{10.3847/1538-4357/ab53d3}

\bibitem[{{Wetzel} \& {Garrison-Kimmel}(2020)}]{GizmoAnalysis}
{Wetzel}, A., \& {Garrison-Kimmel}, S. 2020, {GizmoAnalysis: Read and analyze Gizmo simulations}.
\newblock \doeprint{2002.015}

\bibitem[{{Wetzel} {et~al.}(2023){Wetzel}, {Hayward}, {Sanderson}, {Ma}, {Angl{\'e}s-Alc{\'a}zar}, {Feldmann}, {Chan}, {El-Badry}, {Wheeler}, {Garrison-Kimmel}, {Nikakhtar}, {Panithanpaisal}, {Arora}, {Gurvich}, {Samuel}, {Sameie}, {Pandya}, {Hafen}, {Hummels}, {Loebman}, {Boylan-Kolchin}, {Bullock}, {Faucher-Gigu{\`e}re}, {Kere{\v{s}}}, {Quataert}, \& {Hopkins}}]{Wetzel2023}
{Wetzel}, A., {Hayward}, C.~C., {Sanderson}, R.~E., {et~al.} 2023, \apjs, 265, 44, \dodoi{10.3847/1538-4365/acb99a}

\bibitem[{{Wetzel} {et~al.}(2016){Wetzel}, {Hopkins}, {Kim}, {Faucher-Gigu{\`e}re}, {Kere{\v{s}}}, \& {Quataert}}]{Wetzel2016}
{Wetzel}, A.~R., {Hopkins}, P.~F., {Kim}, J.-h., {et~al.} 2016, \apjl, 827, L23, \dodoi{10.3847/2041-8205/827/2/L23}

\bibitem[{{Wiersma} {et~al.}(2009){Wiersma}, {Schaye}, {Theuns}, {Dalla Vecchia}, \& {Tornatore}}]{Wiersma2009}
{Wiersma}, R. P.~C., {Schaye}, J., {Theuns}, T., {Dalla Vecchia}, C., \& {Tornatore}, L. 2009, \mnras, 399, 574, \dodoi{10.1111/j.1365-2966.2009.15331.x}

\bibitem[{{Willett} {et~al.}(2023){Willett}, {Miglio}, {Mackereth}, {Chiappini}, {Lyttle}, {Elsworth}, {Mosser}, {Khan}, {Anders}, {Casali}, \& {Grisoni}}]{Willet2023}
{Willett}, E., {Miglio}, A., {Mackereth}, J.~T., {et~al.} 2023, \mnras, 526, 2141, \dodoi{10.1093/mnras/stad2374}

\bibitem[{{Wisnioski} {et~al.}(2015){Wisnioski}, {F{\"o}rster Schreiber}, {Wuyts}, {Wuyts}, {Bandara}, {Wilman}, {Genzel}, {Bender}, {Davies}, {Fossati}, {Lang}, {Mendel}, {Beifiori}, {Brammer}, {Chan}, {Fabricius}, {Fudamoto}, {Kulkarni}, {Kurk}, {Lutz}, {Nelson}, {Momcheva}, {Rosario}, {Saglia}, {Seitz}, {Tacconi}, \& {van Dokkum}}]{Wisnioski2015}
{Wisnioski}, E., {F{\"o}rster Schreiber}, N.~M., {Wuyts}, S., {et~al.} 2015, \apj, 799, 209, \dodoi{10.1088/0004-637X/799/2/209}

\bibitem[{{Xiang} \& {Rix}(2022)}]{XiangRix2022}
{Xiang}, M., \& {Rix}, H.-W. 2022, \nat, 603, 599, \dodoi{10.1038/s41586-022-04496-5}

\bibitem[{{Yu} {et~al.}(2023){Yu}, {Bullock}, {Gurvich}, {Hafen}, {Stern}, {Boylan-Kolchin}, {Faucher-Gigu{\`e}re}, {Wetzel}, {Hopkins}, \& {Moreno}}]{Yu2023}
{Yu}, S., {Bullock}, J.~S., {Gurvich}, A.~B., {et~al.} 2023, \mnras, 523, 6220, \dodoi{10.1093/mnras/stad1806}

\bibitem[{{Zinchenko} {et~al.}(2019){Zinchenko}, {Just}, {Pilyugin}, \& {Lara-Lopez}}]{Zinchenko2019}
{Zinchenko}, I.~A., {Just}, A., {Pilyugin}, L.~S., \& {Lara-Lopez}, M.~A. 2019, \aap, 623, A7, \dodoi{10.1051/0004-6361/201834364}

\end{thebibliography}

\appendix
\counterwithin{figure}{section}

\vspace{-6 mm}
\section{Radial growth of all galaxies}
\label{sec:appendix}

\begin{figure*}
\centering
\includegraphics[scale = 0.31]{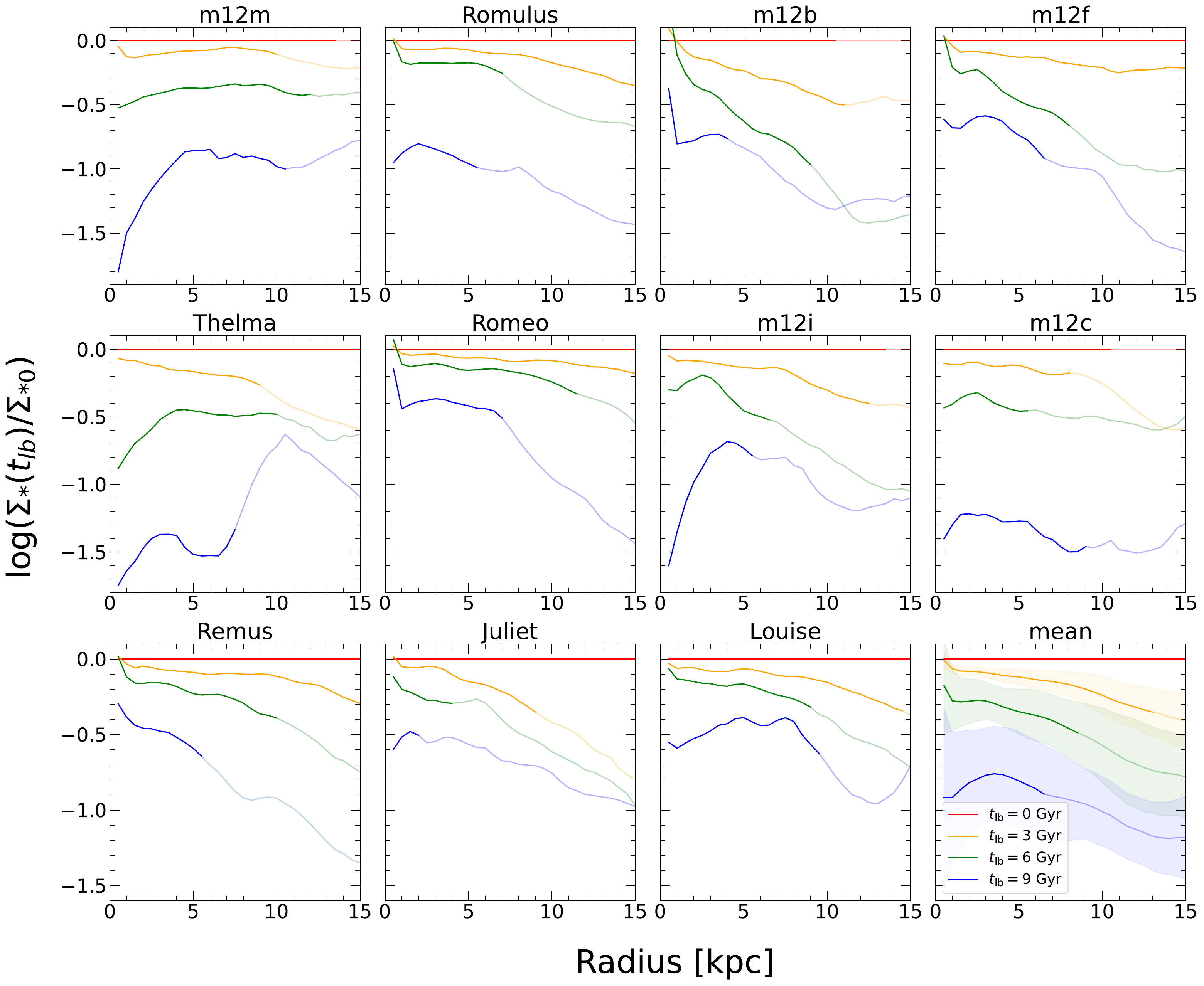}
\vspace{-3 mm}
\caption{
Fractional surface density of all stars at a given lookback time, $t_{\rm lb}$, relative to today, $\log( \Sigma_*(t_{\rm lb}) / \Sigma_{\rm \star,0} )$, versus cylindrical radius in the disk, $R$, for each of the $11$ FIRE-2 galaxies, sorted by decreasing stellar mass today (see Table~\ref{tab:galaxy_properties}).
The bottom right panel shows the mean and standard deviation across these galaxies.
Lines are lighter at $R > R_{\rm \star,young}^{\rm 90}(t_{\rm lb})$.
All galaxies except m12m experienced ``inside-out" radial growth, in terms of greater fractional growth at larger $R$, since at least $t_{\rm lb} = 3 \Gyr$.
$7$ galaxies (m12i, m12f, m12b, Romeo, Romulus, Remus, and Louise) also show inside-out growth since at least $t_{\rm lb} = 6 \Gyr$.
Romeo (the earliest-forming disk) and Remus show inside-out growth since at least $t_{\rm lb} = 9 \Gyr$.
}
\vspace{-5 mm}
\label{fig:all_galaxies}
\end{figure*}

Here we examine the radial growth in each of our $11$ FIRE-2 MW-mass galaxies, using all stars.
Figure~\ref{fig:all_galaxies} shows the ratio of surface density of all stars at a given lookback time to the value today at the same $R$, $\Sigma_\star(t_{\rm lb}) / \Sigma_\star(t_{\rm lb} = 0)$, versus cylindrical radius, $R$, for each galaxy.
The bottom right panel shows the mean and standard deviation across all $11$ galaxies.
Meaningful inside-out growth occurred if the difference in $\log(\Sigma_\star(t_{\rm lb}) / \Sigma_{\star, 0})$ between two lookback times is greater at larger $R$ than at smaller $R$.
All our galaxies experienced inside-out radial growth over at least the last $\approx 3 \Gyr$, except for m12m.
$7$ galaxies (m12i, m12f, m12b, Romeo, Romulus, Remus, and Louise) show inside-out radial growth over the last $\approx 6 \Gyr$.
Only Romeo and Remus show strong inside-out radial growth going back $\approx 9 \Gyr$ ago.
The lookback time of the onset of inside-out growth correlates with when each galaxy started to form a stellar disk \citep[as in][]{McCluskey2023}, such that early (late) disk onset correlates with early (late) onset of inside-out radial growth.
The $3$ latest-forming disks are Juliet ($4.4 \Gyr$ ago), Thelma ($4.4 \Gyr$ ago), and m12c ($6.49 \Gyr$ ago), which do not exhibit inside-out growth until $\approx 3 \Gyr$ ago.
Our $3$ earliest-forming disks, are Romeo ($11.0 \Gyr$ ago), m12m ($9.2 \Gyr$ ago), and Remus ($7.9 \Gyr$ ago), and Romeo and Remus show the earliest onset of inside-out growth.
m12m is a clear exception, which we plan to analyze further in future work.

\end{document}